\documentclass[a4paper,twocolumn,11pt,accepted=2025-07-31]{quantumarticle}
\pdfoutput=1
\usepackage{amsmath,amsfonts,amssymb,amsthm,bbm,graphicx,enumerate,times,mathtools,physics,marvosym,wasysym,soul,stmaryrd,dsfont}
\usepackage{hyperref}
\usepackage[noabbrev,capitalise]{cleveref}
\usepackage[usenames,dvipsnames]{xcolor}
\usepackage{float}
\hypersetup{colorlinks,linkcolor=black,citecolor=ForestGreen,urlcolor=black}
\usepackage[ruled,longend]{algorithm2e}
\usepackage{etoolbox}
\usepackage{comment}

\newcommand{\id}{I} 

\theoremstyle{definition}
\newtheorem{definition}{Definition}

\begin{document}

\title{Far from Perfect: Quantum Error Correction with (Hyperinvariant) Evenbly Codes}

\author{Matthew Steinberg}
\affiliation{QuTech, Delft University of Technology, 2628 CJ Delft, The Netherlands}
\affiliation{Quantum and Computer Engineering Department, Delft University of Technology, 2628 CD Delft, The Netherlands}
\email{matt.steinberg3@gmail.com}
\orcid{0000-0003-3348-7380}

\author{Junyu Fan}
\affiliation{QuTech, Delft University of Technology, 2628 CJ Delft, The Netherlands}
\affiliation{Quantum and Computer Engineering Department, Delft University of Technology, 2628 CD Delft, The Netherlands}
\orcid{0009-0000-3765-2748}

\author{Robert J.~Harris}
\affiliation{ARC Centre for Engineered Quantum Systems, School of Mathematics and Physics, The University of Queensland, St Lucia, QLD, 4072, Australia}
\orcid{0000-0002-8838-6299}

\author{David Elkouss}
\affiliation{QuTech, Delft University of Technology, 2628 CJ Delft, The Netherlands}
\affiliation{Networked Quantum Devices Unit, Okinawa Institute of Science and Technology Graduate University, Okinawa, Japan}
\orcid{0000-0003-2023-2768}

\author{Sebastian Feld}
\affiliation{QuTech, Delft University of Technology, 2628 CJ Delft, The Netherlands}
\affiliation{Quantum and Computer Engineering Department, Delft University of Technology, 2628 CD Delft, The Netherlands}
\orcid{0000-0003-2782-1469}

\author{Alexander Jahn}
\affiliation{Department of Physics, Freie Universit\"at Berlin, 14195 Berlin, Germany}
\email{a.jahn@fu-berlin.de}
\orcid{0000-0002-7142-0059}

\begin{abstract}
We introduce a new class of qubit codes that we call Evenbly codes, building on a previous proposal of hyperinvariant tensor networks.
Its tensor network description consists of local, non-perfect tensors describing CSS codes interspersed with Hadamard gates, placed on a hyperbolic $\{p,q\}$ geometry with even $q\geq 4$, yielding an infinitely large class of subsystem codes.
We construct an example for a $\{5,4\}$ manifold and describe strategies of logical gauge fixing that lead to different rates $k/n$ and distances $d$, which we calculate analytically, finding distances which range from $d=2$ to $d \sim n^{2/3}$. Investigating threshold performance under erasure, depolarizing, and pure Pauli noise channels, we find that the code exhibits a depolarizing noise threshold of about $19.1\%$ in the code-capacity model and $50\%$ for pure Pauli and erasure channels under suitable gauges. We also test a constant-rate version with $k/n = 0.125$, finding excellent error resilience (about $40\%$) under the erasure channel. Recovery rates for these and other settings are studied both under an optimal decoder as well as a more efficient but non-optimal greedy decoder. We also consider generalizations beyond the CSS tensor construction, compute error rates and thresholds for other hyperbolic geometries, and discuss the relationship to holographic bulk/boundary dualities. Our work indicates that Evenbly codes may show promise for practical quantum computing applications.
\end{abstract}

\maketitle

\section{Introduction}

Quantum error correction (QEC) is a necessity for fault-tolerant quantum-information processing at large scale \cite{preskill_nisq}. The utility of QEC codes for practical quantum processing depends on a number of properties, such as finite code rates, a threshold against physical errors, low-weight stabilizer checks, and fault-tolerant protocols for universal logical operations \cite{gottesman_thesis,ft_gottesman,gottesman_qldpc}.
The program of finding specific code constructions with such properties has drawn from a number of fields outside of quantum information theory, such as topological order \cite{topological_quantum_memory,bombin_strong_resilience}, classical coding theory \cite{breuckmann_ldpc}, and more recently, holographic dualities such as the \emph{Anti-de Sitter / Conformal Field Theory} (AdS/CFT) correspondence \cite{happy_code,css_harris,jahn_topical,approx_bacon_shor_code_cao,hyper_mera_cao_pollack,farrelly_tn_codes,harris_phd_thesis}.
In the last approach, one constructs so-called \emph{holographic codes} derived from the bulk-to-boundary encoding maps of model of quantum gravity \cite{Maldacena,witten,klebanov}. Though the bulk degrees of freedom in these models -- which are associated with \emph{logical qubits} of a code -- live in one higher dimension than the boundary, the hyperbolic geometry of the bulk ensures that the Hilbert space of the \emph{physical qubits} on the boundary is larger. The peculiar geometric properties of these maps then imply genuine quantum error correction \cite{bulk_locality_harlow,happy_code}.

Apart from their usefulness in studying bulk / boundary correspondences in theoretical high-energy physics, it has been suggested that holographic codes could have practical application in quantum information science. Indeed, holographic codes based on \emph{absolutely maximally-entangled} (AME) states \cite{max_entangle_states_review,ame_combinatorial_multiunitary} or \emph{planar maximally-entangled} (PME) states \cite{pme_states} (which are also known as \emph{perfect} and \emph{block-perfect} tensors, respectively \cite{css_harris,harris_phd_thesis,farrelly_tn_codes,farrelly_parallel_decoding,perfect_tangles}) have been shown to exhibit several desirable properties for practical quantum error correction. Among these properties are tunable, non-zero rates; high central-logical qubit thresholds for several logical-index implantation schemes (which are comparable to those of several topological codes \cite{topological_quantum_memory,bombin_strong_resilience}); and efficient parallelizable decoders \cite{farrelly_tn_codes,farrelly_parallel_decoding}.

In this work, we explore a recent proposal for \emph{hyperinvariant tensor network} (HTN) codes \cite{steinberg_2023} and show that it encompasses simple qubit-level codes; we dub these new codes \emph{Evenbly codes} after the author of the original HTN paper \cite{evenbly_htn_ansatz}. 
An essential property of these tensor network codes is that they do not require perfect tensors, and can thus be built from gates that mediate less-than-maximal entanglement. 
Our specific construction incorporates seed tensors from $\llbracket q,1,2 \rrbracket_{2}$ CSS codes placed on the vertices of a hyperbolic tiling, along with Hadamard gates on its edges; this can be generalized to any even-$q$ hyperbolic tiling with Schl{\"a}fli symbols $\{ p,q \}$, giving rise to a new, infinite family of holographic subsystem codes. We methodically show that the resulting code structure upholds the isometric constraints typified by the HTN ansatz \cite{conformal_props_steinberg,evenbly_htn_ansatz}, and additionally show how to partition the logical subspace as a subsystem code. The distance and rate of variants of the code is analyzed, with some achieving rates up to $R=\sqrt{3}/2$ and others exhibiting a distance scaling up to $d(n) \sim n^{2/3}$. Moreover, we investigate the erasure threshold of the zero-rate $\{ 5,4\}$ Evenbly code using two types of decoders: a quadratically-scaling greedy reconstruction algorithm and an optimal recovery algorithm based on Gaussian elimination, which exhibits a cubic time complexity \cite{css_harris,optimality_gaussian_alg}. We find that the greedy reconstruction algorithm faithfully provides a lower bound for the threshold (approximately $40\%$), whereas the Gaussian-elimination algorithm obtains a threshold of $49.95 \pm 0.01\%$ in the $Z$ gauge. A constant-rate $k/n = 0.125$ variant of the Evenbly code is tested as well, and it is shown to achieve a threshold of about $44\%$ in the $Y$ gauge for the central logical qubit, and more than $39\%$ in the $Z$ gauge. 
In both the zero- and constant-rate case, the $X$ gauge corresponds to a product state embedding of the ungauged logical qubits on the boundary, preserving the transversal weight-$2$ gates of the $\llbracket 4,1,2 \rrbracket_{2}$ seed code.
We also study the effects of depolarizing and pure Pauli noise on the zero-rate $\{5,4\}$ Evenbly code using the integer-optimization decoder from \cite{harris_int_decoder}, proving the existence of a threshold at $19.1 \pm 0.94\%$ for the former, and around $50\%$ for the latter, ostensibly surpassing the well-known \emph{zero-rate hashing bound} from random coding theory \cite{nielsen_chuang,xzzx_surface_code,wilde_qit}. Other constructions are possible, and we provide several examples that utilize seed tensors which are neither CSS  nor planar 2-uniform, as well as evaluating the erasure thresholds under greedy decoding for several even-$q$ generalizations of the Evenbly code discussed in the main text.

This paper is organized as follows. We start with a synopsis on tensor networks and holographic quantum codes (\cref{section:MERA+HaPPY+HTNs}) and subsequently introduce \emph{hyperinvariant tensor networks} (HTN) as an ansatz (\cref{section:HTNansatz}). Next, we construct Evenbly codes (\cref{section:results}), commencing first with the introduction of 1-uniform (i.e. GHZ) states to the vertex tensors of the original HTN ansatz (\cref{section:HTNGHZ}), and following subsequently with the construction of the qubit-level Evenbly code using \emph{planar 2-uniform states} (\cref{section:htn_codes_from_planar_2_uni_states}). We further show that our construction admits a subsystem structure, in line with previous holographic quantum codes, and we describe how to partition the logical space and perform gauge fixing for the code (\cref{section:gauge_code}). Proceeding, we investigate the rate and distance scaling for the code, and as a result construct a finite-rate version of the Evenbly code (\cref{section:distance}); the threshold properties are investigated for a zero-rate Evenbly code (\cref{section:erasure_threshold}) and a constant-rate variant in \cref{section:constant_rate}; in the process, we explain how the $X$ gauge choice in the Evenbly code allows for weight-2 transversal logical operations to emerge, and we show that very high thresholds emerge in several of the other gauge choices. Finally we report depolarizing and pure Pauli noise threshold results (\cref{section:depo_purepauli_noise}), which are found to be competitive with known code-capacity thresholds for topological codes \cite{bombin_strong_resilience}. The implications of our findings are discussed and concluding comments are provided in \cref{section:conclusion}. In the appendices, we provide additional information on how to build multipartite maximally-entangled (MME) states (\cref{section:appendix_mme_quantum_codes}); an example construction of an Evenbly code using an AME graph state (\cref{section:appendix_htn_code_from_a_graph_state}); Several additional constructions on generalized $\{p,q\}$ tilings and and an accompanying greedy threshold analysis (\cref{section:appendix_css_other_threshold_data}); several seed-tensor constructions that are neither CSS nor planar 2-uniform states, yet still can be used to create new Evenbly codes (\cref{section:appendix_non_css_htn_codes}); and lastly, details on the three different decoder strategies employed in this paper (\cref{section:appendix_decoder_details}).

\section{Background} \label{section:background}

\subsection{Tensor Networks \& Holographic Codes} \label{section:MERA+HaPPY+HTNs}

The first tensor network to be studied as a model for holographic dualities was the \emph{multiscale entanglement-renormalization ansatz} (MERA) \cite{Swingle:2009bg}, originally proposed as an ansatz for approximating groundstates of quantum-critical spin chains \cite{vidal_class_many_body_states_mera,vidal_intro_mera}. Though it exhibits some approximate error-correction features \cite{kastoryano_mera}, its geometry does not match with expectations from AdS/CFT \cite{czech_kinematic_mera,consistency_mera_carroll}.
A more natural tensor network realization of holographic bulk-to-boundary codes was introduced with the \emph{Harlow-Preskill-Pastawski-Yoshida} (HaPPY) codes \cite{happy_code}, a discrete realization of quantum error correction in the AdS/CFT correspondence \cite{bulk_locality_harlow}. The HaPPY code itself can be defined using the \emph{seed tensor} of the ansatz, which determines the encoding isometry for each individual logical qubit, all of which are contracted together as a tensor network on a hyperbolic $\{p,q\}$ tiling.
For the pentagon version ($q=5$) of the HaPPY code, this seed tensor is the encoding isometry for the $\llbracket 5,1,3\rrbracket$ \emph{perfect} stabilizer code \cite{Bennett:1996gf,laflamme_perfect_code}.
This \emph{perfect tensor} also describes a $6$-qubit AME state. The logical qubits of this tensor network code can be recovered using a \emph{greedy algorithm} \cite{happy_code}, so one obtains a concrete dictionary between quantum states in the bulk of the network and boundary states on the uncontracted edges of the tensor network. Other holographic code proposals have also been introduced, mainly in the context of utilizing PME states (such as in \cite{css_harris}) in the bulk rather than AME states.

Holographic codes typically exhibit \emph{uberholography}, i.e., only operators on a fractal subset of the boundary are needed in order to reconstruct logical operators in the bulk \cite{uberholography}. Uberholography is a special case of \emph{operator-algebra quantum-error correction} \cite{lidar_qec_book}, in which a subalgebra representing all possible logical operators in a region is defined using non-local physical boundary operators. HaPPY-style holographic codes have been shown to exhibit high thresholds under both erasure and depolarizing noise models \cite{harris_int_decoder,harris_phd_thesis,farrelly_tn_codes}, as well as exhibiting biased-noise threshold behavior better than the hashing bound \cite{fan2024overcoming,junyu_msc_thesis}. In the spirit of investigating holographic dualities, it has been shown that HaPPY-style holographic codes do not reproduce many of holography's known features at finite $N$ (i.e., under gravitational corrections) \cite{akers2019holographic,dong2019flat}. Such features include non-trivial entanglement spectra of reduced density matrices, state-dependent bulk-reconstruction, and corrections to the entanglement entropy, whose recovery requires tensor network codes with non-perfect tensors \cite{approx_bacon_shor_code_cao,hyper_mera_cao_pollack,steinberg_2023}. 
The quantum code construction that we propose in this paper, in addition to its novel error correction properties, are thus relevant from the perspective of holography as well; we address these attributes in our companion paper \cite{steinberg_2023}.

\subsection{Hyperinvariant Tensor Networks} \label{section:HTNansatz}

\begin{figure*}
\centering
\includegraphics[width=0.9\textwidth]{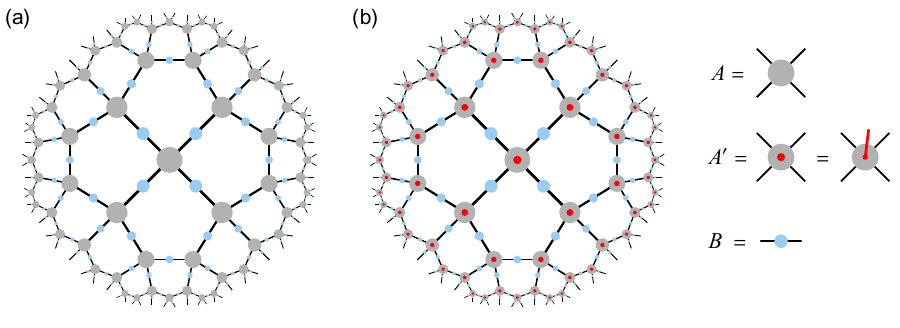}
\caption{Codes from a hyperinvariant tensor network (HTN).
(a) Evenbly's HTN setup from \cite{evenbly_htn_ansatz}, composed of $A$ and $B$ tensors on a hyperbolic $\{p,q\}$ tiling (here for $p=5,q=4$) that respect the symmetries of the infinitely extended tiling.
Open black edges represent a state on the physical boundary qubits, while connected black edges are tensor network contractions. 
(b) The extension of the HTN into a quantum error correction code proposed in \cite{steinberg_2023} (dubbed \emph{Evenbly code}), in which the $q$-leg $A$ tensors are replaced by $q{+}1$-leg $A^\prime$ tensors, with the additional leg (red dot) representing a logical ``bulk'' qubit. The tensor network forms the encoding isometry of this bulk-to-boundary code.
} 
\label{figure:HTN_tilings}
\end{figure*}

\emph{Hyperinvariant tensor networks} (HTN) were proposed for exploring conformal field theories and critical states that are adherent to a description in terms of the AdS/CFT correspondence \cite{evenbly_htn_ansatz,conformal_props_steinberg}. The hyperbolic tessellation and associated bulk hyperbolic symmetries were taken from previous work on holographic codes \cite{happy_code}; however, when combined with the superoperator structure present in MERA \cite{vidal_class_many_body_states_mera,vidal_intro_mera}, the intent was to devise simulations of conformal field theory (CFT) states for which the ansatz upholds a discretized version of conformal symmetry \cite{conformal_quasicrystals,jahn_central_charge,jahn_qcft}. In keeping with the hyperbolic structure characteristic of holographic quantum codes, the HTN ansatz is constructed from a two-dimensional hyperbolic tessellation, as shown in \cref{figure:HTN_tilings}(a); such hyperbolic tilings are defined by Schl{\"a}fli symbols $\{p,q\}$ in the bulk manifold (where $p$ represents the number of sides in a polygon residing in the bulk, and $q$ is the number of edges meeting at each vertex), with the requirement that $\frac{1}{p} + \frac{1}{q} < \frac{1}{2}$ for hyperbolic tessellations. 
Here we follow the notation used in \cite{conformal_props_steinberg,evenbly_htn_ansatz}, which is related to the HaPPY code notation in Ref.\ \cite{happy_code} via a $p \leftrightarrow q$ duality transformation exchanging tiles and vertices. An HTN can be arranged into concentric layers of tensors with which one can perform entanglement renormalization \cite{evenbly_htn_ansatz}. Every concentric layer of the HTN's bulk can be described as realizing a step in the scale-invariant, real-space \emph{renormalization-group} (RG) flow of a typical MERA network. 
There are different prescriptions on how to delineate these concentric layers; here we consider \emph{vertex inflation} \cite{conformal_quasicrystals,jahn_central_charge}, which matches the definition of RG layers in the original HTN proposal \cite{evenbly_htn_ansatz}.

\begin{figure}
\centering
\includegraphics[width=0.8\columnwidth]{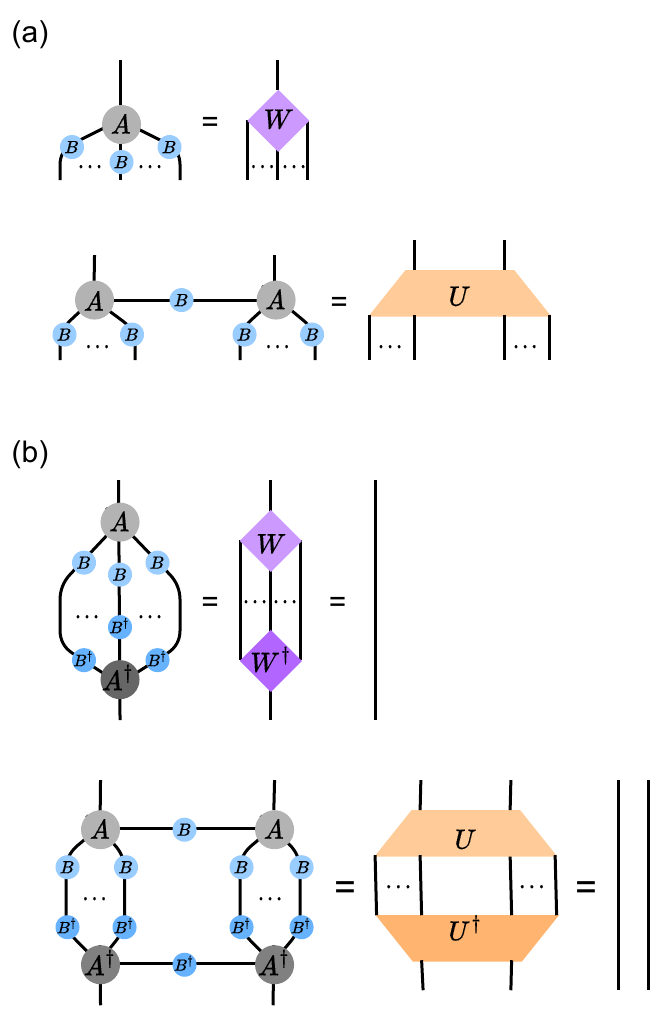}
\caption{Isometry constraints for the HTN defined on a $\{p,q\}$ tiling. We color-code the relevant tensors as follows: light-/dark-gray tensors represent $A$ and $A^{\dagger}$; light-/dark-blue tensors represent $B$ and $B^{\dagger}$; light-/dark-purple tensors depict the $3:1$ isometries $W$ and $W^{\dagger}$; and the $4:2$ isometries $U$ and $U^{\dagger}$ are shown in creme and light orange, respectively.} 
\label{fig:isometry_HTN_general}
\end{figure}

Every vertex and edge in an HTN consists of a set of tensors which we shall call $A$ and $B$, where $A$ corresponds to rotationally-invariant tensors with $q$ indices, and $B$ represents rank-2 rotationally-invariant tensors on every contracted edge of the tessellation. See \cite{evenbly_htn_ansatz,conformal_props_steinberg,steinberg_2023} for detailed information on how to construct an HTN ansatz using a set of isometric tensors $U(A,B)$ and $W(A,B)$. \cref{fig:isometry_HTN_general} also shows an example of isometry constraints which are needed in order to realize entanglement renormalization on the $\{p,q\}$-manifold at every layer, in the spirit of the MERA. The tensors $A$ and $B$ themselves admit various decompositions which fulfill the so-called \emph{isometry constraints} (called \emph{multitensor constraints} in \cite{evenbly_htn_ansatz}). Furthermore, the original HTN ans\"{a}tze employ \emph{doubly-unitary} matrices named $Y,Q$ and $R$ \cite{evenbly_htn_ansatz}. The arbitrariness of $A$- and $B$-tensor decompositions was showcased in \cite{conformal_props_steinberg}, where several other matrices were utilized in order to uphold the isometry constraints.
Such solutions to the HTN isometry constraints come with a set of free parameters, each non-equivalent choice of which leads to a valid HTN solution with different RG flow and correlation decay \cite{conformal_props_steinberg,evenbly_htn_ansatz}, as expected for a quantum-critical system \cite{di_francesco}. 
With a reformulation of solution sets in terms of logical qubits or qudits, one can then construct codes from HTN ans\"atze \cite{steinberg_2023}, with a choice of the logical state determining its free parameters.
Before introducing such a code, however, we begin with a simpler qubit HTN solution that already exhibits some central features. 

\section{Results}\label{section:results}

\subsection{HTN Ans\"{a}tze from 1-Uniform States} \label{section:HTNGHZ}

We begin by constructing first a simpler case of the original HTN ansatz of \cite{evenbly_htn_ansatz}; namely, we prepare a single boundary state rather than a bulk-to-boundary code, which permits us to analyze the tensor network structure more easily. We then show in \cref{section:htn_codes_from_planar_2_uni_states} that this simplified case easily generalizes to a new subclass of holographic quantum codes. 

\begin{definition}\label{theoremHTN}
A hyperinvariant tensor network (HTN) is composed of two tensors $A$ and $B$ which reside on the vertices and edges, respectively, of a $\{p,q\}$ hyperbolic tessellation (with $q$-leg vertices in a $p$-gon tiling).
These tensors comply with the following criteria:
\begin{enumerate}
\item $A$ and $B$ fulfill \emph{isometry constraints} involving tensors on single and neighboring vertices. The constraints are shown explicitly in \cref{fig:isometry_HTN_general} for the $q>3$ case (the special $q=3$ case is covered in \cite{evenbly_htn_ansatz}).
\item $A$ is symmetric under cyclic index permutations, i.e., $A_{k_1,k_2,\dots,k_q} = A_{k_q,k_1,\dots,k_{q-1}}$. 
\item $B$ is a symmetric unitary, i.e., $B=B^\text{T}$ and $B B^\dagger = \id$. 
\end{enumerate}
\end{definition}
The first criterion ensures that the tensor network behaves as an isometric map along any radial direction, while the second and third ensure that the tensors preserve all the (triangle group) symmetries of the  $\{p,q\}$ tessellation (hence the moniker \emph{hyperinvariant}).
These three criteria can be fulfilled by choosing $B = \id$ and $A$ as \emph{perfect tensors} that describe $q$-qudit states that are \emph{absolutely maximally entangled} (AME). This special case of HTNs corresponds to states of HaPPY codes \cite{happy_code}, but also produces flat entanglement spectra for subregions. Avoiding this scenario, Evenbly codes can be restricted further by imposing a fourth criterion:
\begin{enumerate}
    \item[4.] $A$ is chosen as a non-perfect tensor, i.e., there exist bipartitions of its $q$ indices between which it does not mediate maximal entanglement. 
\end{enumerate}
We now introduce a basic example of an HTN that fulfills all four criteria, choosing an $A$ tensor that describes $q$-partite GHZ state and a $B$ tensor that describes a Hadamard gate $H$. For simplicity we will restrict ourselves to qubits, though generalizations to qudits are possible \cite{steinberg_2023}.

Explicitly, the $q$-partite GHZ state vector is defined as 
\begin{equation} \label{eq:EQ_GHZ}
\ket{\text{GHZ}} = \frac{1}{\sqrt{2}} \left[ \ket{0}^{\otimes q} + \ket{1}^{\otimes q} \right] . 
\end{equation}
Equivalently, we can express this state using the stabilizer formalism, where
\begin{subequations}
\begin{align} \label{eq:EQ_GHZ_STAB}
\mathcal{G}_1 &= X^{\otimes q} , \\
\mathcal{G}_{k} &= I^{\otimes k-2} (Z Z) I^{\otimes q-k} \ ,
\end{align}    
\end{subequations}
where $k\in \{2,3,\dots,q\}$. As usual, we end up with $(n-k) = q$ generators for the full stabilizer group. 

GHZ states are 1-uniform \cite{genuine_mme_ortho_arrays,raissi_modify_method} and thus $A$ forms an isometry only from each single site to the remaining ones. For $q>3$, this is a weaker condition than for perfect tensors, which are $\lfloor \frac{q}{2} \rfloor$-uniform, hence fulfilling the fourth criterion. 

Moreover, from \cref{eq:EQ_GHZ}, we can infer rotational invariance as well, thereby satisfying the second criterion. The third criterion is upheld through our choice of $B = H$ with
\begin{equation}\label{eq:EQ_HADAMARD}
H = \frac{1}{\sqrt{2}} 
\begin{bmatrix}
1 & 1 \\
1 & -1 
\end{bmatrix} \ .
\end{equation}

The isometry constraints from the remaining first criterion can be proven by using \emph{operator pushing}: By applying the stabilizers from \cref{eq:EQ_GHZ_STAB}, we can replace a Pauli operator acting on one of the tensor legs by an equivalent operator acting on other legs. Moving Pauli operators past a $B$ tensor involves exchanging $X \leftrightarrow Z$, as $X H = H Z$ and $Z H = H X$.

\begin{figure*}
\centering
\includegraphics[width=\textwidth]{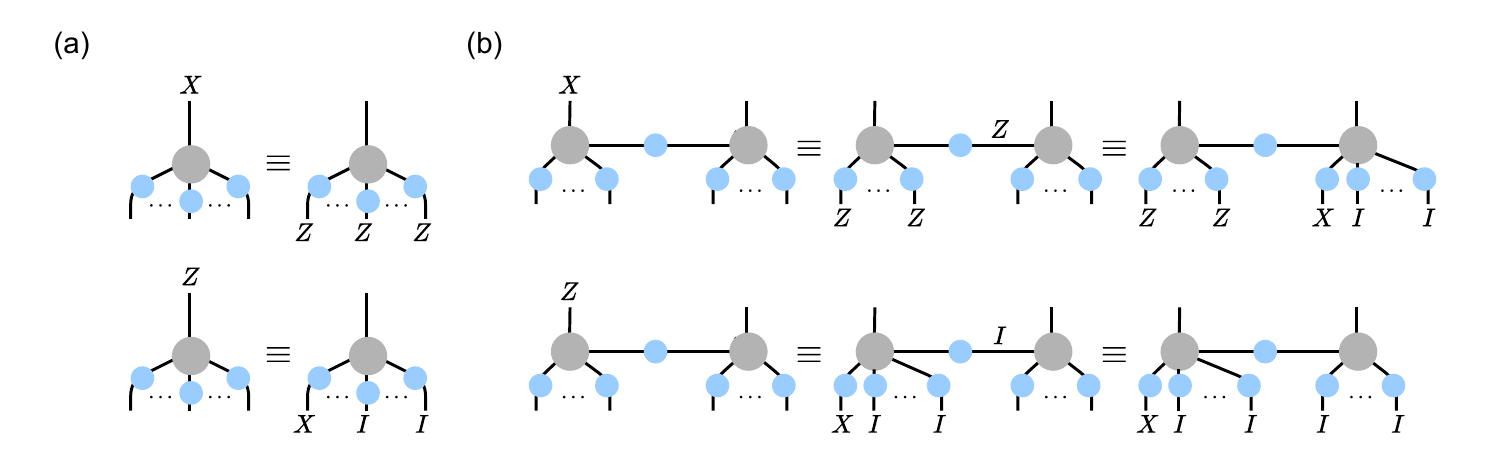}
\caption{Operator recovery on subsystems of the HTN with $A$ tensors (light gray) chosen to represent GHZ states and $B$ tensors (light blue) chosen as Hadamard gates. A) For the isometry $W$, applying the generators from \cref{eq:EQ_GHZ_STAB} removes a Pauli operator on the top leg while adding new ones on the bottom ones. As this is possible for the whole Pauli algebra (generated by X and Z), the tensors form an isometry from top to bottom legs, as expected from the 1-uniformity requirement. Note that pushing through a $B$ tensor swaps $X \leftrightarrow Z$. B) For the $U$ isometry, a Pauli operator can again be pushed to the bottom legs by applying the relation from \cref{eq:EQ_GHZ_STAB}. Due to reflection symmetry, it is sufficient to consider operators acting on only one of the two top legs. These diagrams offer proof that the generalized isometry constraints shown in \cref{fig:isometry_HTN_general} are upheld, thus allowing for the construction of a consummate HTN.}
\label{fig:GHZ_isometry_proof}
\end{figure*}

Showing that each Pauli basis operator X and Z acting on the upper (input) legs of the isometries in  \cref{fig:isometry_HTN_general}(a) can be pushed to a different operator on the lower (output) legs then implies that $W$ or $U$ form an isometry, respectively. The steps of the proof are shown explicitly in \cref{fig:GHZ_isometry_proof}. In these steps, we find that the choice $B=H$ is essential for recovering the X subalgebra in the two-tensor constraint (shown in \cref{fig:GHZ_isometry_proof}(b)), as it converts the problem of removing a single X operator into that of removing a single Z, which can be done by a local generator.

We have therefore found a class of HTNs that provably fulfill the constraints for general $\{p,q\}$, subject to the usual hyperbolic requirement $p>\frac{2q}{q-2}$. One can in principle define an HTN for $q=3$, but in that case our GHZ construction leads to a perfect $A$ tensor, violating the fourth criterion.
Though it is possible to construct a non-perfect Evenbly code by modifying the isometry constraints \cite{evenbly_htn_ansatz}, this case will not be of interest to our construction of Evenbly codes: Associating each bulk vertex with a logical qubit will lead to more bulk than boundary qubits in the case of $\{p,3\}$ tilings, precluding any exact bulk-to-boundary code.

\begin{figure*}
\centering
\includegraphics[width=\textwidth]{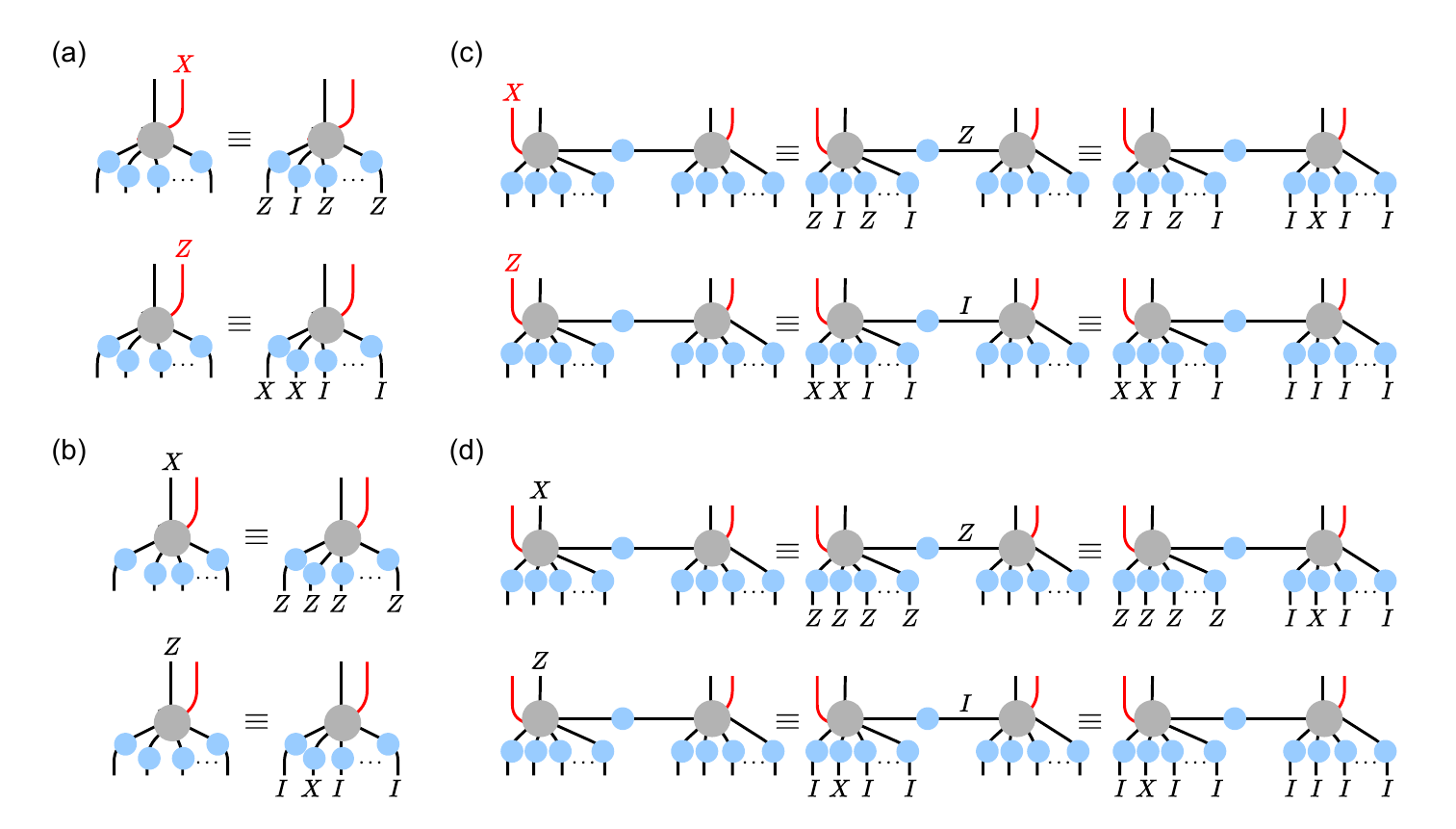}
\caption{Operator recovery on subsystems of the Evenbly code (\cref{eq:EQ_XZCODE_STAB}) via the isometry constraints. A)-B) For a single $A'$ tensor, the logical $\bar{X}$ and $\bar{Z}$ operators (red) can be represented on $\frac{q}{2}$ and $2$ sites, respectively, following \cref{eq:EQ_XZCODE_LOG_OP}. Note that X and Z operators are swapped when pushed through a $B$ tensor. For two adjacent $A'$ tensors and an intermediate $B$ tensor, the $\bar{Z}$ operator on a given logical qubit can be represented on the physical qubits on one side. The $\bar{X}$ operator requires a representation on both sides, with an application of the stabilizers (\cref{eq:EQ_XZCODE_STAB}) removing any Pauli operators acting on contracted legs. By reflection symmetry, the same rules apply for logical operators acting on the second site. C)-D) Any operator acting on a single physical leg can also be pushed to the remaining ones by applying \cref{eq:EQ_XZCODE_STAB}. Together with A)-B), these rules are sufficient to prove the  isometry conditions.}
\label{fig:XZCODE_OPREC}
\end{figure*}

\subsection{HTN qubit codes}\label{section:htn_codes_from_planar_2_uni_states}

\begin{figure}
\includegraphics[width=0.95\columnwidth]{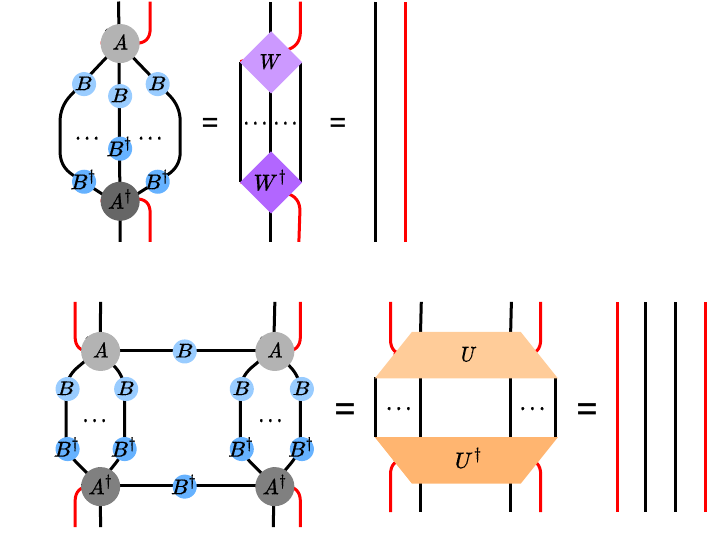}
\caption{Generalized isometry constraints for the Evenbly code. The newly-defined tensor $A'$ in effect describes the complex amplitudes of a $(q+1)$-partite multipartite maximally-entangled state. Red lines denote logical indices.}
\label{fig:HTNcode_isometry_constraints}
\end{figure}

Now we generalize the results gained from \cref{section:HTNGHZ} and define \emph{Evenbly codes}; i.e., HTNs with bulk logical degrees of freedom. Evenbly codes were first defined in \cite{steinberg_2023} as tensor network codes that associate a logical qubit with each vertex in an HTN, a process of \emph{bulk implantation} which replaces each $q$-leg $A$ tensor by a $(q+1)$-leg tensor $A'$. In analogy with \cref{theoremHTN}, an Evenbly code is then defined as follows:

\begin{definition} \label{theoremHTNcodes}
A \emph{Evenbly code} is composed of two tensors $A'$ and $B$, on the vertices and edges, respectively, of a hyperbolic tessellation with Schl\"{a}fli symbol $\{p,q\}$ and $q>3$. These tensors must comply with the following criteria:

\begin{enumerate}
\item $A'$ and $B$ fulfill the \emph{generalized isometry constraints} depicted in \cref{fig:HTNcode_isometry_constraints}.
\item $A'$ defines the encoding isometry for a $\llbracket  q,1,2 \rrbracket$ error-detecting code. 
\item $A'$ is symmetric under cyclic permutations of the planar indices i.e., $A_{j,k_{1},k_{2},\dots,k_{q}} = A_{j,k_{q},k_{1},\dots,k_{q-1}}$, where $j$ denotes the logical index.
\item $B$ is a symmetric unitary matrix, i.e., $B=B^\text{T}$ and $B B^{\dagger} = \mathbb{I}$.
\end{enumerate}
\end{definition}

Note that $A'$ merely describes an error-detecting code with code distance $d=2$; this condition implicitly excludes perfect tensors for $q>4$, which would lead to codes with $d>2$. Though it is possible to define a five-qubit perfect $A^\prime$ at $q=4$, the tensor describing any physical $4$-qubit state after projecting out the logical qubit will generally be non-perfect, thus again leading to tensor network states with non-flat entanglement spectra, generalizing the HTN definition of \cref{section:HTNGHZ}.
Generally, finding solutions to the constraints in \cref{fig:HTNcode_isometry_constraints} is much harder than for those in \cref{fig:isometry_HTN_general} since the isometries $W'$ and $U'$ act on more input sites than $W$ and $U$. 

Ref.~\cite{steinberg_2023} introduced only a single ququart example fulfilling these constraints; we will now show that our GHZ HTN construction can be extended to define a class of qubit codes whose $A^\prime$ tensors define \emph{Calderbank-Steane-Shor} (CSS) codes.
While we focus on this construction in the remainder of the paper, other qubit-level Evenbly codes can be constructed: In \cref{section:appendix_htn_code_from_a_graph_state,section:appendix_non_css_htn_codes}, we introduce Evenbly codes whose seed tensors are not built from CSS codes, with $A^\prime$ tensors describing strictly 2-uniform or AME$(5,2)$ states. 
In contrast, the Evenbly code can be defined for any even $q \geq 4$, and its stabilizer generators defining the $A'$ tensors of rank $(q+1)$ are given by
\begin{subequations}
\label{eq:EQ_XZCODE_STAB}
\begin{align}
\mathcal{G}_1 &= X^{\otimes q} , \\
\mathcal{G}_{k} &= I^{\otimes k-2} (ZIZ) I^{\otimes q-k-1} \ ,
\end{align}
\end{subequations}
where $k \in \{2,3,\dots,q-1\}$. This defines a $\llbracket  q, 1, 2\rrbracket$ code with a logical qubit spanned by the states
\begin{subequations}
\label{eq:EQ_XZCODE_LOG_ST}
\begin{align}
\ket{\bar{0}} &= \frac{1}{\sqrt{2}} \left[ \ket{0}^{\otimes q} + \ket{1}^{\otimes q} \right] \ , \\
\ket{\bar{1}} &= \frac{1}{\sqrt{2}} \left[ ( \ket{0}\ket{1})^{\otimes \frac{q}{2}} + (\ket{1}\ket{0})^{\otimes \frac{q}{2}} \right] \ .
\end{align}
\end{subequations}
Both codeword states are invariant under cyclic permutation of the physical qubits, and hence the encoding isometry represented by $A'$ is invariant under cyclic permutation of the physical indices. This code is closely related to the GHZ state construction with generators in \cref{eq:EQ_GHZ_STAB}, which can be obtained by adding $\bar{Z}$ (defined below) as an additional generator to \cref{eq:EQ_XZCODE_STAB}. In fact, $\ket{\bar{0}}$, $\ket{\bar{1}}$, and any superposition of these two state vectors comprise a GHZ state in different bases, with each single site maximally entangled with the rest of the system. 

The tensor $A'$ is therefore partially 2-uniform when considering the logical and any one physical index, fulfilling the first of the Evenbly code constraints. In fact, it fulfills an even stronger condition: For \emph{any} planar embedding of the logical index, the tensor describes a $q{+}1$-qubit state that is \emph{planar 2-uniform}, i.e., any two qubits that are neighboring under this embedding are maximally entangled with the remainder \cite{planar_kuni_states}. One choice for the logical operators is given by
\begin{subequations}
\label{eq:EQ_XZCODE_LOG_OP}
\begin{align}
\bar{X} &= (IX)^{\otimes \frac{q}{2}} \ , \\
\bar{Z} &= I^{\otimes (q-2)} Z Z \ .
\end{align}
\end{subequations}
Any cyclic permutation of these operators acts equivalently on the code space. Note that this implies that the logical Z subalgebra can be recovered on any two adjacent physical qubits (leading to a code distance $d=2$), while the logical X subalgebra requires $\frac{q}{2}$ non-adjacent sites (for even $q$).

This code defines an $A'$ tensor that forms an Evenbly code if the $B$ tensor is again chosen as the Hadamard gate, as shown in \cref{eq:EQ_HADAMARD}. As we show graphically in \cref{fig:XZCODE_OPREC}, this construction allows us to push any operator acting on the logical or physical \emph{input} legs to the \emph{output} legs of the local tensor(s), analogously to \cref{fig:GHZ_isometry_proof} but with additional logical legs, hence proving the isometry conditions.

Finally, we add that one may use Hadamard gates to exchange the stabilizers and logical operators of this (generalized) code, leading to the form: 
\begin{subequations}
\label{eq:HadamardExchange}
\begin{align}
\mathcal{G}_1 &= Z^{\otimes q} , \\
\mathcal{G}_{k} &= I^{\otimes k-2} (XIX) I^{\otimes q-k-1}, \\
\bar{X} &= I^{\otimes (q-2)} XX \ , \\
\bar{Z} &= (IZ)^{\otimes \frac{q}{2}} \ ,
\end{align}
\end{subequations}
where $k \in \{2,3,\dots,q-1\}$, as before. For the case where the $\{p,4\}$ lattice is bipartite (i.e., the vertices can be assigned two alternating colors), such as for $p=6$, we can therefore absorb all of the $B$ tensors of the HTN by changing every second vertex to represent the alternate code above.

\subsection{Evenbly Codes as Subsystem Codes}\label{section:gauge_code}

Before considering the resilience of Evenbly codes against errors, we first discuss possible gauge restrictions on the bulk logical qubits.
The general setup introduced in \cite{steinberg_2023} and discussed above assumes a logical space composed of all bulk legs, leading to a setting where both the number of logical and physical qubits increase exponentially with the number $L$ of tensor network layers. However, an alternative to this \emph{max-rate} setting is to only consider the logical qubits on a subset of bulk sites, forming a stabilizer \emph{subsystem code} \cite{ft_logical_gates_williamson,poulin_operator_qec} in which the remaining bulk legs define gauge degrees of freedom. 

Let us briefly review subsystem codes. In general quantum error correction codes, one typically divides the physical Hilbert space $\mathcal{H}_{p}$ as $\mathcal{H}_{p} = \mathcal{H} \oplus \mathcal{\bar{H}}$, where $\mathcal{H}$ represents the code subspace, and $\mathcal{\bar{H}}$ is the complement. For more general subsystem codes, it is generally taken that $\mathcal{H}$ can be decomposed as $\mathcal{H} = \mathcal{H}_{L} \otimes \mathcal{H}_{G}$ \cite{poulin_operator_qec}, where $\mathcal{H}_{L}$ and $\mathcal{H}_{G}$ represent the \emph{logical} and \emph{gauge} subsystems of the code subspace, respectively. 
In this paradigm, logical information is encoded into $\mathcal{H}_{L}$, while the subsystem $\mathcal{H}_{G}$ is used to help diagnose and correct errors in $\mathcal{H}_{L}$. This aim can be accomplished by defining a \emph{gauge group} $\mathcal{G}$, of which the total stabilizer group $\mathcal{S}$ is the center of $\mathcal{G}$, i.e. $\{iI,\mathcal{S}\} = \mathcal{Z}(\mathcal{G}) := C(\mathcal{G}) \cap \mathcal{G}$. Here $iI$, $\mathcal{Z}(\cdot)$, and $C(\cdot)$ represent the identity operator (up to a phase), the center, and the centralizer, respectively \cite{lidar_qec_book}. Up to phase factors, therefore, elements from $\mathcal{G}$ are either stabilizer operators (and act trivially on $\mathcal{H}_{L} \otimes \mathcal{H}_{G}$) or operators which act non-trivially on $\mathcal{H}_{G}$. Lastly, due to the subsystem structure, two types of logical operators arise. \emph{Bare logical operators} are those which act non-trivially only on $\mathcal{H}_{L}$, while \emph{dressed logical operators} act non-trivially on $\mathcal{H}_{L} \otimes \mathcal{H}_{G}$ \cite{ft_logical_gates_williamson,bombin_gauge_fixing}.

\begin{figure*}
\centering
\includegraphics[width=\textwidth]{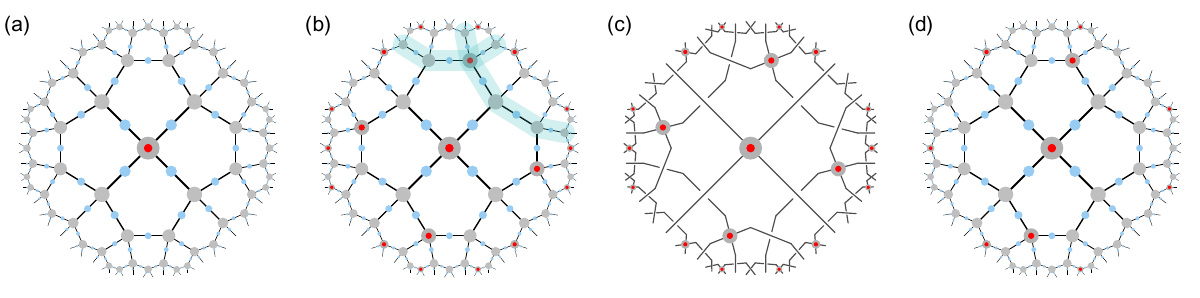}
\caption{Gauged Evenbly codes considered in this paper.
(a) A zero-rate Evenbly code where only the central logical qubit (red dot) is considered, all other bulk legs being gauge-fixed (gray disks).
(b) A constant-rate Evenbly code, where each logical qubit occupies a unique crossing point in the bulk. Here, hyperbolic geodesics only pass over sites that are gauge-fixed. An example for such a pair of geodesics for one logical qubit is shaded in blue.
(c) In the $X$ gauge, all the gauge-fixed tensors act as pairs of overlapping EPR pairs, mapping each logical qubit to four unique physical boundary sites under the initial $\llbracket 4,1,2 \rrbracket$ code. Any other gauge mixes the encodings of logical qubits.
(d) A variant of the code in (b), with half of all logical qubits on every layer gauge-fixed. We examine this code's threshold under erasure in \cref{section:constant_rate}.
}
\label{fig:45_gauge_fixing}
\end{figure*}

In addition to the max-rate setting of Evenbly codes, we consider two subsystem code settings: First, a \emph{zero-rate} case where all logical information is encoded in the central logical qubit and all other logical legs form the gauge subspace $\mathcal{H}_{G}$ (\cref{fig:45_gauge_fixing}(a)). Second, a \emph{constant-rate} case where we only include a fraction of logical qubits on each layer, chosen so that the hyperbolic geodesics crossing over the logical site do not pass over any other ungauged sites. The $\{5,4\}$ example of this second case is visualized in \cref{fig:45_gauge_fixing}(d).

We show below that this constant-rate version of the Evenbly code yields a rate of $R = k/n = 1/8$; for comparison, the \emph{trivial gauge subsystem} variant (i.e. the maximum number of bulk logical indices are used for storing logical information and gauge group is empty) of the $\{5,4\}$ Evenbly code exhibits an asymptotic rate of $R = 0.866$. Such finite encoding rates are a common feature of codes defined on hyperbolic manifolds \cite{hyperbolic_surface_code1,happy_code}.
Both of the code variants'  threshold properties will be evaluated in \cref{section:constant_rate}. In both cases, $\mathcal{H}_{\text{bulk}}$ refers to all other logical qubits partitioned into the gauge subsystem.

Finally, we discuss how to perform \emph{gauge fixing} for Evenbly codes. The concept of gauge fixing originates in \cite{paetznick_gauge_fixing,bombin_gauge_fixing}; here, the concept amounts to adding an element $g \in \mathcal{G}$ into the stabilizer group $\mathcal{S}$ and in parallel removing the element $h \in \mathcal{G}$ (which anticommutes with $g$). For Evenbly codes, we define three ways to gauge-fix $\mathcal{H}_{G}$: we designate these gauges either the \emph{X}, \emph{Y}, and \emph{Z} gauges. In order to form the gauge-fixed stabilizer group $S'$, we simply add logical operators from $\mathcal{H}_{G}$ in accordance with the appropriate logical eigenstate that we wish. For example, forming the $X$ gauge-fixed stabilizer group can be done by simply defining $\mathcal{S}'_{\bar{X}} := \mathcal{S} \cap \mathcal{G}_{\bar{X}}$, where $\mathcal{G}_{\bar{X}}$ represents the X operators acting on $\mathcal{H}_{G}$. Similarly, one can form any logical-eigenstate gauge in our picture simply by adding the appropriate logical operators as $\mathcal{S}'_{\bar{P}} := \mathcal{S} \cap \mathcal{G}_{\bar{P}}$, where $\mathcal{P}$ is an element of the Pauli group. 

\subsection{Rate and Distance Scaling} \label{section:distance}

Calculating the asymptotic rate of a $\{5,4\}$ Evenbly code can be done by following the prescriptions found in \cite{harris_phd_thesis,happy_code,jahn_central_charge}, and is related to an inflation procedure for hyperbolic tilings known as \emph{vertex inflation} \cite{conformal_quasicrystals,jahn_central_charge}. As in these previous examples, rates can be evaluated by comparing the number of tensors in the bulk with the total number of boundary sites. 

Under vertex inflation, one defines each layer of the tiling as a closed loop of vertices (gray disks in our figures), with the initial (zeroth) layer consisting only of the central tensor. 
One then distinguishes the vertices, each of which hosts a logical qubit, by the number of outgoing planar legs pointing from each vertex directly towards the next layer; at the last layer, these open legs support the physical qubits. 
We denote vertices with one to four outgoing legs with $\alpha,\beta,\gamma$ and $\delta$.
The letter $\delta$ only appears as the central (seed) tensor with four planar legs, and $\gamma$ never appears at all.
The sequence of vertices on subsequent layers can be found by applying the substitution rules
\begin{subequations}
\begin{align}
    \alpha &\mapsto \alpha\beta \ , \\
    \beta &\mapsto \alpha\beta\beta\alpha\beta \ , \\
    \delta &\mapsto \alpha\beta\beta\alpha\beta\beta\alpha\beta\beta\alpha\beta\beta \ .    
\end{align}
\end{subequations}
For example, applying two inflation steps to the central tensor produces the sequences
\begin{align}
  \delta \mapsto (\alpha\beta\beta)^4 \mapsto (\alpha\beta\alpha\beta\beta\alpha\beta\alpha\beta\beta\alpha\beta)^4 \ ,
\end{align}
where we have used powers to indicate repetitions of the sequence. This corresponds to the tilings shown in \cref{figure:HTN_tilings} and \cref{fig:45_gauge_fixing}: We find that the first and second layer around the central vertex indeed contain $12$ and $48$ vertices, respectively, with the latter consisting of $n_\alpha=20,n_\beta=28$ vertices of each type. It follows that this tensor network has $n=n_\alpha + 2n_\beta=76$ physical boundary sites.
More formally, vertex inflation determines a \emph{substitution matrix} $M$; every column of $M$ represents the number of sites of each type that arise in every subsequent layer as a result of applying the inflation rules. Combining this matrix with the vector $\eta=(n_\alpha,n_\beta,n_\delta)^\text{T}$ which represents the number of sites at the $L^{\text{th}}$ layer, we construct a recursion relation of the form
\begin{align}
\eta^{(L+1)} \equiv 
\begin{pmatrix}
n_\alpha^{(L+1)} \\
n_\beta^{(L+1)} \\
n_\delta^{(L+1)}
\end{pmatrix} &= 
\begin{pmatrix}
1 & 2 & 4 \\
1 & 3 & 8 \\
0 & 0 & 0
\end{pmatrix}
\begin{pmatrix}
n_\alpha^{(L)} \\
n_\beta^{(L)} \\
n_\delta^{(L)}
\end{pmatrix} \nonumber\\
&\equiv M \eta^{(L)} \ .
\end{align}
We can use our initial vector $\eta^{(0)} = (0 \; 0 \; 1)^\text{T}$ in order to recursively calculate the number of logical sites at each layer via
\begin{equation}
\eta^{(L)} = M^L \eta^{(0)} \ ,
\end{equation}
and compute the numbers $n$ and $k$ of physical and logical qubits as 
\begin{align}
    n^{(L)} &= n_\alpha^{(L)} + 2 n_\beta^{(L)} + 4 n_\delta^{(L)} \ , \\
    k^{(L)} &= \sum_{i=0}^L (n_\alpha^{(L)} + n_\beta^{(L)} + n_\delta^{(L)}) \ .
\end{align}
Given our initial vector $\eta^{(0)}$ and the substitution matrix $M$ for the $\{5,4\}$ code, we can explicitly resolve these expressions in terms of the nonzero eigenvalues $\lambda_1 = 2 + \sqrt{3}$ and $\lambda_2 = 2 - \sqrt{3}$ of $M$, leading to 
\begin{align}
    n_\alpha^{(L)} &=  (2\sqrt{3}-2) \lambda_1^L - (2\sqrt{3}+2) \lambda_2^L  \ , \\
    n_\beta^{(L)} &= 2 \lambda_1^L + 2 \lambda_2^L \ , \\
    \label{EQ_54_N_PHYSICAL}
    n^{(L)} &= (2\sqrt{3}+2) \lambda_1^L - (2\sqrt{3}-2)\lambda_2^L \ ,  \\
    k^{(L)} &= 1 + \frac{2\sqrt{3}(\lambda_2^{L+1}+\lambda_2^{-L}+\sqrt{3}-3)}{\sqrt{3}-1} \ .
\end{align}
for $L>0$. Despite appearances, all of these evaluate as integers.
The rate of the code follows as
\begin{equation}
    R^{(L)} \equiv \frac{n^{(L)}}{k^{(L)}} = \frac{5-5\sqrt{3} + 2\sqrt{3}( \lambda_2^{L+1} + \lambda_2^{-L} )}{4(\lambda_1^L - (2-\sqrt{3}) \lambda_2^L ) }\ ,
\end{equation}
which quickly converges to the asymptotic rate
\begin{equation}
\bar{R} \equiv \lim_{L \to \infty} \frac{k^{(L)}}{n^{(L)}} = \frac{\sqrt{3}}{2} \approx 0.866 \ .
\end{equation}
This is the rate of the max-rate Evenbly code in which all logical qubits are gauge-free. However, in \cref{section:constant_rate}, we treat another variant, where a number of logical qubits are gauge-fixed. 
The symmetries of the $\{5,4\}$ tiling allow for a particularly elegant gauge choice: Each vertex in the tiling is an orthogonal crossing point of two geodesics through the hyperbolic disk (shown in \cref{fig:45_gauge_fixing}(c)), with the new geodesics at each layer passing either over a series $\beta-\beta$ of two vertices or over $\beta-\alpha-\beta$ of three. Choosing only every second $\beta$ vertex at each layer as the location of a gauge-free logical qubit then associates each vertex with its ``own'' pair of geodesics than cross no other gauge-free logical qubit.
As each geodesic ends in two physical sites on the boundary, it is unsurprising to find that this code has the constant rate
\begin{align}
    R^\prime = \frac{\frac{1}{2}\sum_{i=0}^L n_\beta^{(L)}}{ n_\alpha^{(L)} + 2 n_\beta^{(L)} + 4 n_\delta^{(L)}} = \frac{1}{4} \ .
\end{align}
In our threshold studies we will consider an even sparser version of this model, where only every fourth $\beta$ vertex remains gauge-free (\cref{fig:45_gauge_fixing}(d)), leaving the rate $R' = 1/8$.

We now consider code distances, i.e., the minimum physical weight of any logical operator, distinguishing between \emph{bit distances} $d_\text{bit}$ of single-qubit logical operators in the bulk basis and \emph{word distances} $d_\text{word}$ of arbitrary multi-qubit logical operators. 
In most literature, $d_\text{word}$ is simply referred to as the \emph{distance} $d$ of the code \cite{knill2000theory}.
For the Evenbly code, both bit and word distances heavily depend on the choice of gauge: In the max-rate case, acting with two physical Pauli operators can create a logical operation arbitrarily deep in the bulk, leading to $d_\text{word}=2$. If the two Pauli operators act on neighboring physical qubits, they will affect only logical qubits adjacent to the boundary (e.g.\ $Z_j Z_{j+1}$ acting as a $\bar{Z}$ on a single tile); however, if they act further apart, the result is a string of logical operators along a shortest path throughout the bulk, a feature that is related to the appearance of non-trivial 2-point functions in such codes \cite{steinberg_2023}.

However, the bit distance of logical operators $\bar{X}_i, \bar{Z}_i$ on a particular logical qubit on the $i$th tile generally increases exponentially with its distance to the boundary. 
For the max-rate $\{5,4\}$ code, we can compute this bit distance by again utilizing inflation rules, focusing on the logical qubit in the center of the hyperbolic tiling. Applying the operator pushing rules for the stabilizer generators $( XXXX, IZIZ, ZIZI )$, we find the following behavior under vertex inflation: An operator X acting on the boundary of the layer $L$ turns into a Z on layer $L+1$, while a Z turns into three operators $Z,X,Z$ acting on different boundary qubits. It is this second step which results in an exponential growth of Pauli weight with $L$, and we can compute this growth exactly. Denoting the number of Xs and Zs of a given logical operator on the $L$th layer as $w_X^{(L)}$ and $w_Z^{(L)}$, we arrive at the rule
\begin{align}
    \begin{pmatrix}
    w_X^{(L+1)} \\
    w_Z^{(L+1)} 
    \end{pmatrix} =
    \begin{pmatrix}
    0 & 1 \\
    1 & 2 
    \end{pmatrix} 
    \begin{pmatrix}
    w_X^{(L)} \\
    w_Z^{(L)} 
    \end{pmatrix}~,
\end{align}
where the substitution matrix is obtained from the mapping $X \mapsto Z$ and $Z \mapsto ZXZ$, as described above. At the central tile, we begin with $(w_X^{(0)},w_Z^{(0)}) = (2,0)$ for $\bar{X}$ and $(0,2)$ for $\bar{Z}$. From this it follows that the total Pauli weight $w^{(L)}=w_X^{(L)} + W_Z^{(L)}$ on a given layer $L$ is
\begin{subequations}
\begin{align}
    w^{(L)}(\bar{X}) &= (1 + \sqrt{2})^L  + (1 - \sqrt{2})^L \ , \\
    w^{(L)}(\bar{Z}) &= (1 + \sqrt{2})^{L+1}  + (1 - \sqrt{2})^{L+1} \ .
\end{align}    
\end{subequations}
Note that the weight of $\bar{Z}$ is exactly one inflation step ahead of $\bar{X}$; hence, the bit distance is given by $d_\text{bit}= w^{(L)}(\bar{X})$.
It is generally desirable to express the scaling of $d_\text{bit}$ in terms of the number $n$ of physical qubits, which we computed in \cref{EQ_54_N_PHYSICAL}. At large $L$, the scaling becomes
\begin{align}
\label{EQ_DIST_SCALING_MAX_RATE}
    d_\text{bit} \sim (1 + \sqrt{2})^L &= \left(\frac{n}{2 + 2\sqrt{3}}\right)^{\log_{2+\sqrt{3}} (1+\sqrt{2})}  \nonumber\\
    &\approx 0.32\, n^{0.67} \ .
\end{align}
The distance for $\bar{Z}$ scales similarly, with an additional prefactor of $1+\sqrt{2}$. However, in general the distance of the entire code is upper-bounded as $d_{\text{word}} \leq d_{\text{bit}}$. To achieve a more useful code, we would like to increase $d_\text{word}$ (towards the upper bound dictated by $d_{\text{bit}}$) for a greater resilience against errors so that targeted logical operations are easier to perform. 

This condition is exactly what gauge-fixing some logical qubits of the Evenbly code achieves: In the most extreme example of the zero-rate code with only a single gauge-free logical qubit in the center of the tiling, $d_\text{word} = d_\text{bit}$ while being upper-bounded by the max-rate result from \cref{EQ_DIST_SCALING_MAX_RATE}. This upper bound may not be tight, as the availability of additional stabilizer terms on gauge-fixed tiles allows for more compact representations of logical operators on the boundary. Exactly computing these distances is a computationally difficult problem, though we expect that gauge-fixing in a Z or $Y$ gauge will reduce the power coefficient in \cref{EQ_DIST_SCALING_MAX_RATE} somewhat below $0.67$. 
Note that the arguments for a $d_\text{bit} = O(n^{0.67})$ bound should apply to a logical qubit at any bulk location as layers are added, though prefactors will be highly depend on its distance to the boundary; for example, a logical qubit on the last layer (assumed not to be gauge-fixed) will always have weight-$2$ logical operators. 

A special case is given by the zero-rate code in the $X$ gauge, where the weight of logical operators does not scale at all with the number of layers, resulting in $d_\text{word} = d_\text{bit} = 2$. This result also extends to the constant-rate code of \cref{fig:45_gauge_fixing}(c) and its sparsifications, e.g.\ the rate $R^\prime = 1/8$ code in the $X$ gauge. As a result, it can be easily shown using operator-pushing rules that the transversal logical operations of the seed tensor (i.e. $\bar{X}, \bar{Z}, \overline{CX}$) are of constant weight-2 in this gauge. 

\subsection{Erasure Threshold for a Zero-Rate Code} \label{section:erasure_threshold}

We analyze the threshold performance of the $\{5,4\}$ Evenbly code using two different decoding strategies for erasure errors: A \emph{greedy reconstruction} method generalizing the approach of \cite{happy_code} adapted to our GHZ-based Evenbly code, in which the bulk logical qubits are iteratively reconstructed from the boundary and whose threshold we denote by $p^{\text{greedy}}_{\text{erasure}}$, and a \emph{Gaussian elimination} algorithm adapted from \cite{css_harris} with threshold $p^{\text{gaussian}}_{\text{erasure}}$. 
The implementation and simulation details for both algorithms can be found in \cref{section:appendix_decoder_details}. To summarize, greedy reconstruction relies on simple geometric steps but is less effective than the Gaussian elimination approach, which considers all possible physical representations of logical operators.

\begin{figure*}
\centering
\includegraphics[width=\textwidth]{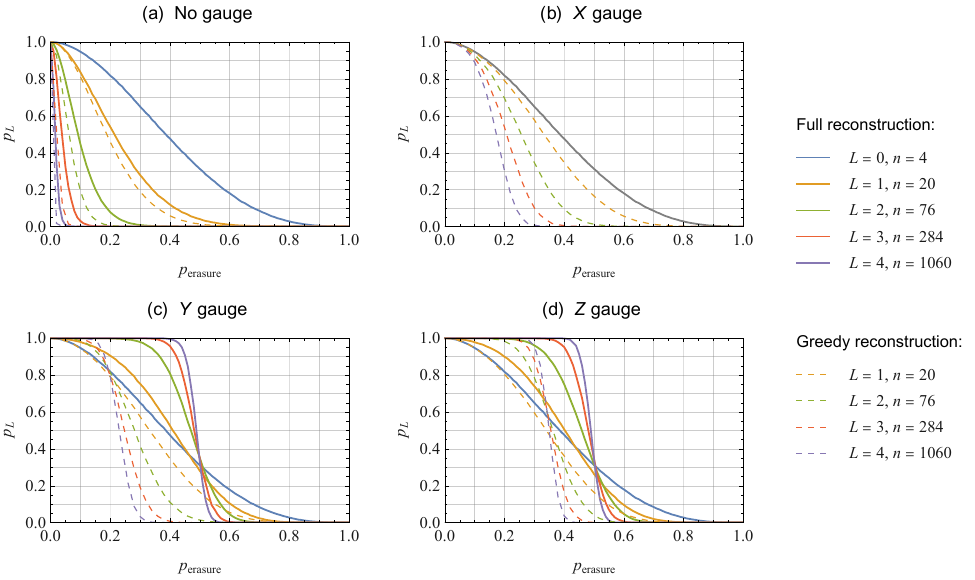}
\caption{Recovery probabilities $p_L$ for the $\{5,4\}$ Evenbly code under an erasure noise model with noise rate $p_\text{erasure}$. 
We consider recovery of logical operators under a greedy algorithm (dashed curves) and full algebraic recovery via the Gaussian elimination algorithm of \cite{harris_phd_thesis,css_harris}, with details on the implementation in \cref{section:appendix_decoder_details}.
(a) No threshold appears for the ungauged (max-rate) case. 
(b) In the logical $X$ gauge of the zero-rate code, logical operators have the same support at every scale length. (c-d) In the logical Y and $Z$ gauges, we observe high resilience against erasures, obtaining a threshold of $49.83 \pm 0.09\%$ and $49.95 \pm 0.01\%$ in either case, respectively. Moreover, the greedy reconstruction decoder obtains thresholds in the range of approximately $24\%$ and $40\%$, for the Y and $Z$ gauges, respectively.}
\label{figure:rate_zero_erasure}
\end{figure*}

Simulation results for recovery rates with both decoding algorithms under an erasure channel are shown in \cref{figure:rate_zero_erasure}, considering only the recovery of the central logical qubit.
In subfigures \ref{figure:rate_zero_erasure}(a)-(d), the threshold results $p_{\text{erasure}}$ are shown for each choice of gauge fixing discussed in \cref{section:gauge_code}. 
No gauge fixing (subfigure \ref{figure:rate_zero_erasure}(a)) corresponds to the max-rate case, whereas subfigures \ref{figure:rate_zero_erasure}(b)-(d) consider different gauges of the zero-rate code. In agreement with the discussion of a ququart Evenbly code in \cite{steinberg_2023}, the max-rate case is highly susceptible to erasure errors and does not display a threshold.

This property is a feature of the holographic construction, in which non-vanishing two-point functions for all boundary states are desirable. In the zero-rate case for logical $X$ gauge (subfigure (b)), recovery probabilities for the full recovery algorithms are independent of the number $L$ of tensor network layers. Under this gauge, $A^\prime$ corresponds to a state of two EPR pairs, each between the two opposite ends of the four sites of the tensor. The initial $\llbracket 4,1,2 \rrbracket$ code is thus directly mapped onto four boundary qubits (up to local Hadamard gates), with the remaining boundary qubits forming sets of disconnected EPR pairs. 
Even though the code properties do not change, we find a weaker greedy reconstruction at higher $L$, whose local reconstruction steps becomes less effective as the relevant physical qubits move further apart on the boundary.

The logical $Y$ and $Z$ gauges (subfigures 7(c)-(d)) are different: Here we find a threshold close to the maximal value of $50\%$ for full reconstruction, with the simpler greedy reconstruction only achieving around $24\%$ and $40\%$, respectively. This relative underperformance of the greedy algorithm is due to its reconstruction of the bulk algebra on edges tile-by-tile, which may be prevented by erasure errors even though the logical operators of the central qubit are still recoverable. For this reason, the greedy recovery probability always lower-bounds full recovery, as already seen in other work \cite{harris_phd_thesis,css_harris}.
Conversely, the Gaussian elimination algorithm poses an upper bound to the recovery probability, as it has been shown to be optimal for erasure noise \cite{css_harris,optimality_gaussian_alg}. 

\subsection{Erasure Threshold for a Constant-Rate Code} \label{section:constant_rate}

As shown in \cref{fig:45_gauge_fixing}(b), one may also gauge-fix some of the logical qubits while keeping the rate of the code finite, leading to the constant-rate setting described in \cref{section:gauge_code}. 
Keeping the two graph geodesics across any logical qubit gauge-fixed on all other bulk sites, we potentially increase the protection of each logical qubit from erasures closest to it on the boundary. In the $X$ gauge, this behavior directly generalizes the zero-rate case: Again, the gauged sites turn into EPR pairs that directly map a $\llbracket 4,1,2 \rrbracket$ code onto four unique physical qubits, but with $k=n/4$ logical qubits instead of one. This is visualized in \cref{fig:45_gauge_fixing}(c).

\begin{figure*}
\centering
\includegraphics[width=\textwidth]{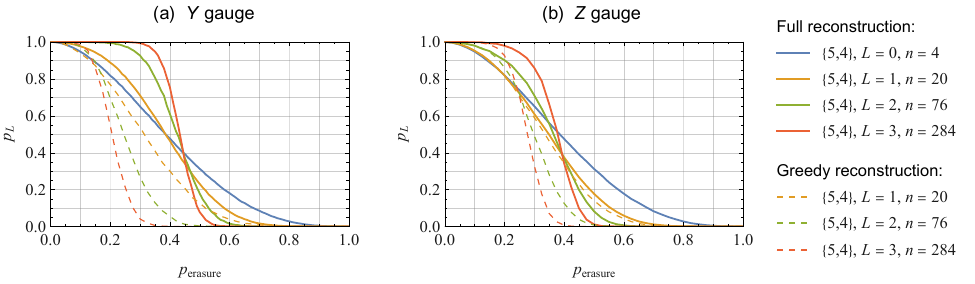}
\caption{Recovery probabilities $p_L$ for a constant-rate $\{5,4\}$ Evenbly code under an erasure noise model with noise rate $p_\text{erasure}$. We consider the rate $1/8$ code shown in Fig.\ \ref{fig:45_gauge_fixing}(d), with recovery of logical operators on the central logical qubit under a greedy algorithm (dashed curves) and full algebraic recovery via the Gaussian elimination algorithm. The gauged bulk qubits are set to either (a) $Y$ gauge or (b) $Z$ gauge, the former showing a slightly higher full threshold but a lower greedy threshold. As for the zero-rate code in Fig.\ \ref{figure:rate_zero_erasure}(b), an $X$ gauge leads to trivial scaling with the number of layers $L$ and is therefore omitted here. 
}
\label{figure:rate_const_erasure}
\end{figure*}

The case of either the $Y$ or $Z$ gauge again leads to non-trivial $L$ dependence and an asymptotic threshold, as in the zero-rate case; the recovery curves are shown in Fig.\ \ref{figure:rate_const_erasure}. They can be boosted by adding completely $Y$ gauge-fixed layers of tensors to the constant-rate code, pushing them close to $50\%$ while decreasing the rate accordingly, as we report in \cref{table:constant_rate_thresholds}.
Note that for such codes, a threshold for the central logical qubit implies a threshold for all the others, as the threshold in a holographic code depends on how much each tensor network layer suppresses errors \cite{css_harris,harris_int_decoder,farrelly_tn_codes,happy_code}. 

Curves for $p_\text{erasure}$ thus become sharper as one considers logical qubits deeper in the bulk. As shown in \cite{farrelly_parallel_decoding}, in holographic codes the recovery of non-central logical qubits can be parallelized as long as the physical error rate stays below the threshold of the central logical qubit.

\begin{table}
\centering
{\scriptsize
\begin{tabular}{|c | c | c | c|} 
 \hline
Variant & $p^{\text{gaussian}}_{L_{Y}}$ & $p^{\text{gaussian}}_{L_{Z}}$ & $R = k/n$ \\ [0.5ex] 
 \hline
0 extra layers  & $44.02 \pm 2.89\%$ & $39.24 \pm 0.6\%$ & $0.125$ \\ 
1 extra layer & $46.61 \pm 0.03\%$ & $43.05 \pm 1.2\%$ & $0.0335$ \\
2 extra layers & $47.80 \pm 0.2\%$ & $45.00 \pm 1.0\%$ & $0.009$ \\ [1ex] 
 \hline
\end{tabular}
}
\caption{Estimated thresholds $p_{\text{erasure}}$ for the central logical qubit, using the Gaussian elimination algorithm for the constant-rate variant of the $\{5,4\}$ Evenbly code. The $Y$ gauge results outperform those of Z, even as more layers of gauged protection are added at the boundary. As we continue adding layers of protection to the code, we asymptotically approach the $50\%$ threshold; however, the rate $R$ of the code also decays to zero in this limit.}
\label{table:constant_rate_thresholds}
\end{table}

\subsection{Threshold for Depolarizing and Pure Pauli Noise} \label{section:depo_purepauli_noise}

\begin{figure*}
\centering
\includegraphics[width=\textwidth]{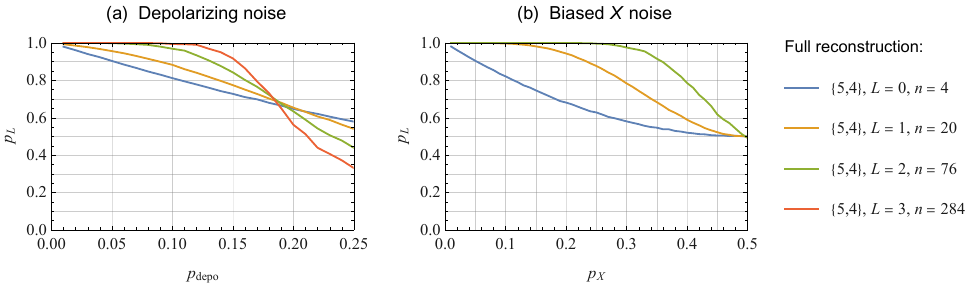}
\caption{Recovery curves for the zero-rate $\{5,4\}$ Evenbly code for depolarizing and biased Pauli noise, under $Z$ gauge fixing. 
(a) A threshold appears at approximately $19.1 \pm 0.94\%$ for depolarizing noise.
(b) For pure Pauli $X$ noise, a preliminary threshold appears at approximately $50\%$. The same behavior was observed for pure Pauli $Y$ and $Z$ noise as well.
}
\label{fig:depo_pure_pauli}
\end{figure*} 

Finally, we investigate here the threshold performance of the zero-rate $\{5,4\}$ Evenbly code for depolarizing noise. The depolarizing noise model can be written succinctly as 
\begin{equation}
\mathcal{E}(\rho) = (1-p)\rho + p/3 \big( \sum_{i}\mathcal{P}_{i}\rho \mathcal{P}_{i} \big),
\end{equation}
in which it is \emph{not assumed} that knowledge of the error's locations are known, in contrast to erasure noise \cite{linear_erasure1,linear_erasure2,harris_int_decoder}. Here, we utilize the \emph{integer-optimization decoder} from \cite{harris_int_decoder,harris_phd_thesis}, which is known to be optimal (although scaling exponentially badly). More details on the simulation can be found in \cref{section:appendix_depo_noise_details}. 

In \cref{fig:depo_pure_pauli}, (a) displays the recovery threshold $p_{\text{depo}}$ obtained as a result of varying the physical noise parameter. As is shown, a threshold appears in the region of approximately $p_{\text{depo}} \approx 19.1 \pm 0.94\%$; this threshold value is very competitive with known code-capacity thresholds for the surface code, color code, and the holographic $\llbracket 6,1,3 \rrbracket$ code \cite{bombin_strong_resilience,farrelly_tn_codes,xzzx_surface_code}. Subfigure 9(b) displays the results obtained for pure Pauli-X noise using the integer-optimization decoder (we have additionally tested pure Pauli-Y and Pauli-Z noise, and have found the same results). As can be ascertained, preliminary estimates indicate pure Pauli-X noise thresholds of approximately $50\%$, in line with the results obtained for the erasure channel. However, as we were limited in the computational power and time available, we report these findings here as preliminary, with three crossing points. However, if confirmed, it would seem that the Evenbly code is capable of reaching and exceeding known achievable bounds for random zero-rate quantum codes (known as the zero-rate \emph{hashing bound} \cite{wilde_qit,xzzx_surface_code}).  

\section{Discussion} \label{section:conclusion}

\subsection{qCFTs \& Holography}

The construction presented in this work is one among many possible realizations of Evenbly codes; indeed, it was shown in \cite{conformal_props_steinberg} that many possible tensor decompositions for the original HTN ansatz exist, even those which are non-unitary in terms of individual tensor properties. However, constructing a quantum error-correcting code with non-unitary tensors may prove to be very challenging, and even more so when trying to determine relevant error correction properties of the resulting code. 

Although we have realized an exact holographic code which exhibits genuine polynomially-decaying boundary-to-boundary correlation functions, we have yet to investigate whether CFT data relevant for a known minimal-model CFT \cite{di_francesco}, or a quasiperiodically disordered qCFT \cite{jahn_qcft}, can appear as a result of the gauge-fixing techniques described. Indeed, recent work \cite{bistron2025bulk} a MERA-style tensor network was proposed in order to capture certain aspect of discretized AdS/CFT, in addition to recent efforts to parameterize and classify central charges of qCFTs based on their tiling structure \cite{jahn_central_charge}. We leave this issue for future work.

Evenbly codes at their core circumvent a no-go theorem mentioned in \cite{hyper_mera_cao_pollack} on the existence of exact holographic codes with polynomially-decaying correlation functions. Recent work suggests that a bulk-to-boundary code replicating full quantum gravity should be expressed in terms of an approximate error-correcting code \cite{kelly2017,faulkner2020holographic,continuous_symmetries_preskill,akers_minimal_surfaces,hayden_alpha_bits}, and that stabilizer codes in particular only exhibit trivial \emph{area operators} \cite{stabilizer_trivial_cao}, meaning that quantum superpositions of bulk geometries on subregions are restricted to classical values for the area of the subregion boundaries (Ryu-Takayanagi surfaces). As discussed in \cite{steinberg_2023}, this still allows for restricted quantum gravity effects where the bulk subregion accessible from a boundary subregion can fluctuate between configurations with equal area.

An interesting implication of our work is that gauge-fixing, which in the holographic picture corresponds to a choice of a given local (quantum) geometry, implies different reconstruction properties and thresholds. For example, the $X$ gauge in the $\{5,4\}$ Evenbly code leads to $A$ tensors describing 4-qubit PME states composed of EPR pairs, resembling the ``fixed-area states'' described by HaPPY codes \cite{akers2019holographic,dong2019flat}. Different gauge choices can thus be seen as a modification of bulk entanglement to suppress or enhance certain types of operators under a bulk RG flow. 
It is therefore possible that understanding the suitability of holographic codes under different types of errors will also lead to new insights into how quantum gravity emerges in such tensor network models.

\subsection{Error Correction}

The results conveyed demonstrate that Evenbly codes achieve similar logical recovery rates when compared to both topological and \emph{quantum low-density parity-check} (qLDPC) codes \cite{breuckmann_ldpc,bombin_strong_resilience} in the code-capacity regime, as well as other holographic codes \cite{farrelly_tn_codes,farrelly_parallel_decoding,harris_int_decoder,css_harris,happy_code}. 
The rate of an Evenbly code is also tunable, a property generally ascribed to holographic codes \cite{happy_code,harris_phd_thesis}.
However, in contrast to the surface or color codes that can achieve a distance scaling $d=O(\sqrt{n})$ for a single logical qubit \cite{lidar_qec_book}, the $\{5,4\}$ Evenbly code exhibits an $O(n^{0.67})$ scaling. 
This distance scaling can be subtle for Evenbly codes with multiple logical qubits, however: Though the weight of logical operators of a single logical bulk qubit (i.e., its bit distance $d_\text{bit}$) still generally scales as $O(n^{0.67})$ for a $\{5,4\}$ code, it is highly dependent on the bulk location and can be as low as $d_\text{bit}=2$.
Giving more precise details on what types of lower and upper bounds can be expected for given orientation patterns of bulk logical qubits in Evenbly codes is in general still an open problem, and is the subject of ongoing research.
Another important extension along this direction would be the examination of \emph{phenomenological noise models} (which incorporate measurement errors); we leave this idea for future work. 

It is known that the seed tensors of our construction, $k$-uniform states, exhibit generally less entanglement than their AME and PME counterparts used in other constructions of holographic codes \cite{max_entangle_states_review}. As the task of generating an AME state is itself considered a benchmark for quantum-processor performance \cite{cerveralierta_ame_state_circuits,nikiforos_matt_ame_paper}, $k$-uniform states are more suitable for experimental implementation since they can be prepared using local interactions only, as recent experiments have shown \cite{wallraff_andersen2020,kuni_prep_cluster_states_experiment}. This observation, together with recent advances in non-Euclidean lattice architectures \cite{kollar} as well as in the preparation of holographic logical code states \cite{huber_engineering_holography} underscores the potential for experimental feasibility, particularly in architectures with flexible qubit connectivity, such as trapped-ion, neutral-atom, or photonic devices \cite{photonics_review,trapped_ion_review,photonics_review2,neutral_atom_review}.

Additionally, we reported that the recovery rate for the central logical qubit in the $R = 1/8$ Evenbly code is $44.02 \pm 2.89\%$ in the $Y$ gauge. At first glance, this result may violate the \emph{quantum channel-capacity theorem} \cite{wilde_qit}, which sets a strict upper bound for erasure channel capacity at $43.75\%$ in our case. However, the channel capacity obtained is well within the error bounds in our simulation, and it is known that as long as physical error rates are below the threshold of the central logical qubit, all other bulk logical qubits can be decoded in parallel in holographic codes \cite{farrelly_parallel_decoding}. In this way, the central logical qubit's recovery rate dictates the practical upper bound for all other bulk logical qubits. Our work therefore suggests that the word recovery rate for a constant-rate Evenbly code is only slightly lower than that of the central logical qubit, and may achieve channel capacity for finite-rate quantum erasure channels. In upcoming work, we determine conclusively whether or not this is indeed the case for the entire class of holographic quantum codes with finite rate \cite{alex_logical_recovery}. 

In addition to the code construction presented in the main text, we have also explored the threshold properties of other examples on $\{p,q\}$ hyperbolic tilings using the greedy recovery algorithm. These results are shown in \cref{section:appendix_css_other_threshold_data}, and depict very interesting (if not similar) threshold data to the example fashioned in the main text. As stated in \cref{section:appendix_htn_code_from_a_graph_state,section:appendix_non_css_htn_codes}, several other non-CSS seed-tensor constructions using AME and 2-uniform quantum states can be utilized to build an Evenbly code. These alternative Evenbly codes may be difficult to decode using the standard greedy decoder, owing to the lack of rotational invariance of the seed tensors. However, the Gaussian elimination or integer-optimization algorithms employed in this work could be used there; one may also consider modifying existing tensor-network decoders \cite{farrelly_parallel_decoding,farrelly_tn_codes} to accommodate Hadamard gates on the edges of the Evenbly code, as well as memory usage due to the exponentially-scaling number of sites observed in vertex inflation \cite{conformal_quasicrystals}.

As a final comment in this section, no prior work to our knowledge has yet considered \emph{fault-tolerant protocols} \cite{ft_gottesman} for a holographic code, although fault-tolerant preparation of precursor $k$-uniform states has recently been studied \cite{majidy2025scalable}. Ref.\ \cite{cao_prd_thresholds} discovered the existence of a threshold in continuum AdS/CFT models, arising as a confinement/deconfinement phase transition. As the majority of holographic codes are stabilizer codes, one can minimally expect a generic holographic code to benefit from the implementation of a fault-tolerant syndrome-extraction protocol such as the \emph{flag-style} protocols of \cite{ft_chamberland_flag,bhatnagar2023low}. However, more elegant solutions may be possible, such as those from \cite{cao2025growing}; although not exhibiting bounded-weight check operators like in quantum low-density parity-check (qLDPC) codes \cite{breuckmann_ldpc}, one may expect holographic codes to be amenable to techniques similar to those from recent work demonstrating excellent space and time overhead savings in the context of generalized concatenated codes \cite{yamasaki1,yamasaki2,goto2024many}. As providing for fault-tolerant operation of the code is a vital step towards the experimental realm, this topic is the subject of active research. 

\subsection{Logical Gate Sets}

A no-go theorem regarding locality-preserving gates outside of the Clifford group for holographic codes was recently presented in \cite{ft_logical_gates_williamson}. Here, holographic codes based on AME and PME states (i.e. approximately upholding the principle of subsystem complementarity) were shown to not admit locality-preserving logical operations outside of the Clifford group. These results were extended to holographic models for which complementary recovery is not exactly fulfilled on any entanglement wedge. Naively, one would ordinarily rule out non-Clifford locality-preserving logical operations for the Evenbly code as well. However, we suspect that the set of locality-preserving logical gates for Evenbly codes may be gauge-dependent, and significantly larger than simply the set known for the $\llbracket q,1,2\rrbracket_{2}$ seed tensor, since the unique, mutable isometry constraints of our construction were not taken into account in \cite{ft_logical_gates_williamson}. This fact is significant since we have shown that new non-local bulk reconstruction rules emerge and substantially alter the threshold properties of the Evenbly code, in accordance with the appropriate gauge selection. It is known that gauge fixing can be used in certain topological code constructions in order to obtain universal fault-tolerant logical gate sets \cite{bombin_gauge_fixing,paetznick_gauge_fixing}, and such a picture may arise in our codes as well. Also, it was shown in \cite{cao_quantum_lego} that holographic codes with transversal logical gates outside of the Clifford group can easily be constructed, taking the form of a \emph{holographic quantum Reed-Muller code}. In any case, a rigorous examination of the fault-tolerant logical gate set for Evenbly codes will be the subject of future work.

Interestingly, it was also mentioned in \cite{cao_quantum_lego} that the quantum Reed-Muller code's seed tensor exhibits 2-uniformity, equivalent to seed tensors that we present for alternative Evenbly codes in \cref{section:appendix_non_css_htn_codes}. Constructing an Evenbly code from quantum Reed-Muller seed tensors and investigating the resulting error correction properties could also prove fruitful. Lastly, one can consider \emph{heterogeneous holographic constructions} with the Evenbly code, in which layers or seed tensors per layer are alternated in some pattern; we report on this in upcoming work \cite{heterogeneous_holo_qrm} and will address the subject of constructing holographic codes with large, potentially universal logical gate sets more thoroughly.

In this article, we have generalized the HTN ansatz to qubit-level codes, which we dub \emph{Evenbly codes}. We have shown that this new, infinitely large code class exhibits unique and rich error correction properties that we have analyzed and reported on. Some of the highlights of this code include: high erasure, depolarizing, and pure Pauli noise thresholds; low-weight transversal logical operators in the $X$ gauge; excellent distance scaling for bulk logical qubits; and constant-rate variants which are easy to tune and contrive. Lastly, our construction utilizes seed tensors which have already been prepared in the laboratory, implying that this class of holographic code may be amenable for experimental realization. \\

\section*{Software Availability}

A portion of this work was completed using an upcoming open-source python package \cite{fan2024lego_hqec} for the simulation of holographic quantum codes and partially uses functionality from \cite{hypertiling}. All of the thresholds determined using the Gaussian elimination and integer-optimization decoders were obtained by utilizing the DelftBlue supercomputer \cite{DHPC2022}. For this study, the Gurobi optimization package \cite{gurobi} was also used for calculating the depolarizing noise thresholds. 

\section*{Acknowledgements}

We thank Aritra Sarkar, Charles Cao, and Jens Eisert for useful discussions. MS and SF thank the Intel corporation for financial support. AJ received funding from the Einstein Research Unit ``Perspectives of a quantum digital transformation''. RH was supported by the Australian Research Council Centre of Excellence for Engineered Quantum Systems (Grant No. CE 170100009). DE was supported by the JST Moonshot R\&D program under Grant JPMJMS226C.  

\section*{Author Contributions}

MS and AJ developed the theoretical constructions and formalism for Evenbly codes. AJ developed and implemented the greedy reconstruction algorithm. JF and MS constructed the Gaussian elimination algorithm. JF constructed and implemented the integer-optimization decoder during his master thesis, and was supervised by MS and SF. RH provided guidance for writing the Gaussian elimination and the integer-optimization decoders. MS and AJ wrote the manuscript. DE, SF, and AJ supervised and coordinated the project, as well as providing insight and guidance during the writing process. 



\appendix

\section{Multipartite Maximally-Entangled States \& Quantum Codes} \label{section:appendix_mme_quantum_codes}

MME states are highly-entangled many-body states in which certain reductions of the state are maximally mixed \cite{max_entangle_states_review,facchi2008maximally}; such states have found numerous applications in the construction of quantum error correction codes and quantum-communications protocols, among others \cite{nielsen_chuang,ame_combinatorial_multiunitary,cerveralierta_ame_state_circuits}. In the case of systems consisting of two qubits, the Bell state $\ket{\Psi^{+}_{2}} = \frac{1}{\sqrt{2}}(\ket{00}+\ket{11})$ is well-known as the simplest quantum state which is maximally entangled, since the state obtained after performing a partial trace on either of the two subsystems results in a maximally-mixed quantum state $\text{tr}_{A}\big[ \ket{\Psi^{+}_{2}}\bra{\Psi^{+}_{2}}\big] = \mathbb{I}_{2}/2$. This state is locally equivalent to other states of the form $(U_{A} \otimes U_{B})\ket{\Psi_{2}^{+}}$, where $U_{A},U_{B} \in U(2)$. In a similar way, one can define a generalized Bell state for any bipartite quantum system with $n$ levels:
\begin{equation}
\ket{\Psi^{+}_{n}} = \frac{1}{\sqrt{n}} \sum_{i=1}^{n} \ket{i}_{A} \otimes \ket{i}_{B}.
\end{equation}
Here, one can easily check that the resulting RDMs for subsystem $A$ or $B$ will indeed be maximally mixed, as the entanglement entropy $\mathcal{S}$ of such states can be expressed as $\mathcal{S}(\ket{\Psi^{+}_{n}}\ket{\Psi^{+}_{n}}) = \log{n}$.

As one increases the number of constituent subsystems to three or more subsystems, the process of characterizing quantum entanglement becomes more complicated as well, with several measures currently in use \cite{max_entangle_states_review}. In this manuscript, we will consider maximal entanglement to be defined by considering the possible bipartitions of a many-body quantum state and the reduced state's corresponding entanglement entropy. In what follows, we will give several examples of maximally-mixed states using the aforementioned definition, which will be useful for the present work. 

In vein with the notion of bipartite maximally-mixed reductions, \emph{$k$-uniform} states are defined as pure states of $n$ distinguishable qudits in which all k-qudit reductions of the entire system are maximally mixed \cite{kuni_mixed_states,genuine_mme_ortho_arrays,raissi_modify_method,raissi_constructions_beyond_mds_codes,planar_kuni_states}, with $k \leq \lfloor \frac{n}{2} \rfloor$. It is well-known that $k$-uniform states themselves are dual descriptions of stabilizer codes \cite{raissi_modify_method,raissi_constructions_beyond_mds_codes}, and that several inequivalent methods for constructing $k$-uniform states exist \cite{genuine_mme_ortho_arrays,raissi_constructions_beyond_mds_codes,raissi_modify_method}. More specifically, a $k$-uniform state $\ket{\psi_{\text{$k$-uniform}}}$ in a Hilbert space $\mathcal{H}(n,q) \in \mathbb{C}^{\otimes n}_{\chi}$ (where $\chi$ represents the local Hilbert-space dimension of all $n$ qudits), denoted by $k$-uniform${(n,q)}$ exists when the following condition holds: 
\begin{equation}
\text{tr}_{\mathcal{\bar{A}}}\big[ \ket{\psi_{\text{$k$-uniform}}}\bra{\psi_{\text{$k$-uniform}}} \big] \propto \mathbb{I}, 
\end{equation}
where $\mathcal{\bar{A}}$ represents the set complementary to $\mathcal{A}$, for all $\mathcal{A} \subset \{1, \cdots , n\}, |\mathcal{A}| \leq k$. The Schmidt decomposition shows that values of $k$ generally are bounded as $k \leq \lfloor \frac{n}{2}\rfloor$, with the condition $k = \lfloor \frac{n}{2}\rfloor$ known as the \emph{absolutely maximally-entangled} (AME) state condition \cite{max_entangle_states_review,ame_combinatorial_multiunitary,cerveralierta_ame_state_circuits} (or equivalently \emph{perfect tensors} \cite{happy_code,jahn_majoranadimers}), and the case of 1-uniform states is commonly associated with the famous \emph{Greenberger–Horne–Zeilinger} (GHZ) state \cite{nielsen_chuang}. Another notion of maximal entanglement was introduced in \cite{pme_states}, in which $k = \lfloor \frac{n}{2}\rfloor$ reductions are necessary, but only for $\lfloor \frac{n}{2}\rfloor$ \emph{adjacent} parties; these quantum states are known as \emph{planar maximally-entangled} (PME) states (or equivalently the \emph{block-perfect tensors} of \cite{harris_phd_thesis,css_harris,jahn_majoranadimers}). Yet another type of maximal entanglement has recently been shown to exist and is known as the so-called \emph{planar $k$-uniform} states \cite{planar_kuni_states}.

Although several methods currently exist for constructing $k$-uniform states, we shall focus on the techniques utilized in \cite{kuni_mixed_states,genuine_mme_ortho_arrays} for the sake of presenting a coherent formalism. In particular, we employ on the construction of $k$-uniform states from \cite{kuni_mixed_states}, using particular sets of $n$-qubit Pauli operators, as a restriction of the stabilizer formalism that is already well-known \cite{nielsen_chuang,gottesman_thesis}, together with the fact that any state can be written as a tensor product of Pauli operators \cite{kuni_mixed_states}. Recalling the form of $n$-qubit Pauli operators, we use the convention 
\begin{align}
\sigma_{X} \otimes \sigma_{Y} \cdots \sigma_{Z} \otimes \sigma_{I} \equiv XY \cdots ZI,
\end{align}
where $\sigma_{j}, \forall j \in \{I,X,Y,Z\}$ are the qubit-level Pauli operators. In accordance with the stabilizer formalism, we can consider a set of these $n$-qubit Pauli operators $\mathcal{G}$
\begin{align}
\mathcal{G} = \{\mathcal{G}_{1}, \cdots, \mathcal{G}_{m}\},
\end{align}
such that these $m$ operators have the following three properties:

\begin{itemize}
\item[$\circ$] Commutation: every pair of elements $\mathcal{G}_{i}, \mathcal{G}_{j}$ must commute, i.e. $[\mathcal{G}_{i},\mathcal{G}_{j}] = 0, \forall i,j \in \{1,...,m\}$, and 
\item[$\circ$] Independence: $\mathcal{G}^{j_{1}}_{1} \cdots \mathcal{G}^{j_{m}}_{m} \propto \mathbb{I} \text{ iff } j_{1} = j_{2} = \cdots = j_{m} = 0$.
\item[$\circ$] $k$-uniformity: the product $\mathcal{G}^{j_{1}}_{1} \cdots \mathcal{G}^{j_{m}}_{m}$ must result in an $n$-qubit Pauli operator containing no more than $(n-k-1)$ Pauli $I$ operators. 
\end{itemize}

By summing up all possible products of generators, one can write the density matrix of a $k$-uniform state as:
\begin{align} \label{equation:rhos}
\rho = \frac{1}{2^{n}}\sum^{1}_{j_{1}\cdots j_{m} = 0}\mathcal{G}_{1}^{j_{1}} \cdots \mathcal{G}_{m}^{j_{m}}.
\end{align} 
This construction follows from the fact that any arbitrary quantum state can be written down using Pauli matrices and a \emph{correlation tensor}, as utilized in \emph{quantum state tomography} \cite{kuni_mixed_states,nielsen_chuang}.

It is additionally well-known that pure $k$-uniform states themselves form a set of quantum codes, formed as superpositions of classical \emph{maximum distance-separable} (MDS) codes \cite{raissi_modify_method,raissi_constructions_beyond_mds_codes}. A particular class of quantum codes known as \emph{stabilizer quantum codes} can be modified via a technique known as \emph{shortening}, and one can show the existence of a pure quantum stabilizer code whose spanning vectors are $(k-1)$-UNI states. Such codes can be defined for any local-dimension Hilbert space, and correspond to qudit-based stabilizer codes.

An $n$-qubit quantum stabilizer code which encodes $k$ \emph{logical qubits}, is represented with the notation $\llbracket n,k,d\rrbracket_{\chi}$ (where $\chi$ represents the local dimension for each of the $n$ particles in the $n$-partite quantum system), and the distance $d$ represents the Hamming-weight difference between two \emph{codewords} defined in a quantum code. The rate of such a code is $R = k/n$. 

The \emph{stabilizer} $\mathcal{S}$ of a code is an \emph{Abelian subgroup} of the $n$-fold Pauli group $\prod^{n}$, and does not contain the operator $I^{\otimes n}$; the simultaneous $+1$ eigenspace of operators forms what is known as the \emph{codespace}. A stabilizer code exhibits a minimal representation of its stabilizer $\mathcal{S}$ in terms of $(n-k)$ independent generators:
\begin{align}
\{\mathcal{G}_{1}, \cdots ,\mathcal{G}_{(n-k)} | \forall i \in \{1, \cdots ,(n-k)\}, \mathcal{G}_{i} \in \mathcal{S}\}.
\end{align}
The generators of the stabilizer function in the same way as the \emph{parity-check matrix} for classical linear codes \cite{macwilliams_sloane_classical_ec}. Moreover, stabilizer generators commute with each other, and errors are generally identified and diagnosed by noting anticommutation with errors that may occur \cite{gottesman_thesis,nielsen_chuang}.

Quantum error correction codes are capable of correcting a number of errors, determined by their distance $d$, and represented by the following equation
\begin{align}
t = \left \lfloor \frac{d-1}{2} \right \rfloor,
\end{align}
where $t$ is the \emph{Hamming weight} of the error.

\section{An Evenbly code from a Graph State} \label{section:appendix_htn_code_from_a_graph_state}

As a first attempt to devise a specific example of an Evenbly code, we start by utilizing the HTN defined on the $\{5,4\}$ tiling from \cite{conformal_props_steinberg,evenbly_htn_ansatz}; we do not consider the $\{7,3\}$ (i.e. $q > 3$ as a basic requirement) simply because the conditions for bulk reconstruction prohibit recovery of bulk logical qubits \cite{steinberg_2023}. 

Rather than considering the ququart construction introduced in previous work \cite{steinberg_2023}, it is in fact possible to construct qubit-level Evenbly codes. We utilize a vertex tensor $A'$ corresponding to a 2-uniform state, which is given by any state in the ground space of the $\llbracket 5,1,3\rrbracket$ perfect code \cite{laflamme_perfect_code,genuine_mme_ortho_arrays}.

\begin{figure*}
\centering
\includegraphics[width=0.95\textwidth]{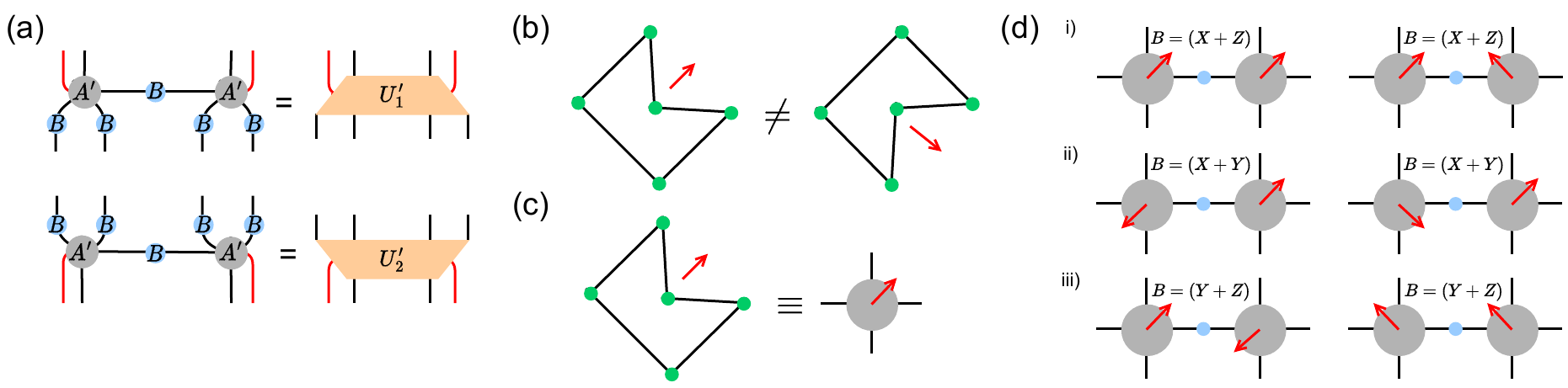}
\caption{A $\{5,4\}$ Evenbly qubit-level code. (a) Complying with the  constraints requires that any pair of neighboring tensors $A'$ with an intermediate tensor $B$ must contract to unitaries $U'_{1}$ and $U'_{2}$ along two directions. (b)-(c) As described in \cref{eq:EQ_4QUBIT_CODE}, the tensor $A'$ is built from the amplitudes of the 5-qubit graph state, shortened into a $\llbracket 4,1,2\rrbracket$ code; in this graph-theoretic form, vertices (in green) denote qubits initialized as $\ket{+}$, and edges represent two-qubit CZ gates. As it is known that this quantum code is not invariant under planar index rotations, the tensors $A'$ have an inherent directionality, denoted by a black arrow.
(d)) Depending on the relative orientations of neighboring $A'-B-A'$ tensors concurrent in a given layer, one of three $B$ tensors is needed to fulfill the conditions in A), here shown up a normalization $\frac{1}{\sqrt{2}}$.}
\label{fig:graphcode}
\end{figure*}

For convenience, we choose the physical state encoding the $\ket{\bar{+}}=\frac{\ket{\bar{0}} + \ket{\bar{1}}}{\sqrt{2}}$ logical state, which happens to coincide with the (AME) 5-qubit cycle graph state \cite{eisert_graph_states1,helwig_ame_graph_states}. This logical state is the unique ground state of the stabilizer Hamiltonian
\begin{align}
H_{\llbracket  5,0,3\rrbracket} = - &\left( ZXZII\right. + IZXZI + IIZXZ \nonumber\\
&+ \left. ZIIZX + XZIIZ \right) \ ,
\end{align}
using the stabilizer generator $S_1= ZXZII$ and its cyclic product permutations. Through code shortening \cite{gottesman_thesis}, we convert this $\llbracket 5,0,3\rrbracket$ state into a  $\llbracket 4,1,2\rrbracket$ code (see \cref{fig:graphcode}B)-C) with stabilizer Hamiltonian
\begin{align}
\label{eq:EQ_4QUBIT_CODE}
    H_{\llbracket  4,1,2 \rrbracket} = -\left(  ZXZI + IZXZ + XZZX \right) \ .
\end{align}
The logical operators of this code can be represented as $\bar{X}=ZIIZ$ and $\bar{Z}= IIZX$. Unlike the original HTN construction \cite{evenbly_htn_ansatz,conformal_props_steinberg} and the ququart Evenbly code \cite{steinberg_2023}, this code is not symmetric under cyclic permutation of physical indices. Therefore, a suitable $B$ tensor to satisfy the isometry constraints has to be found for each of the 16 possible rotations (6 up to reflections) of two neighboring $A'$ tensors with a $B$ in between; consequently, the constraints allow for bulk reconstruction from any direction. For any pair of neighboring $A'$ tensors, this implies that the $B$ tensor connecting them has to fulfill two isometry constraints, shown in \cref{fig:graphcode}A.

We find that for the $A'$ tensor following from \cref{eq:EQ_4QUBIT_CODE}, three different $B$ tensors are necessary to fulfill the constraints for various rotations, visualized in \cref{fig:graphcode}D. These three tensors are the Hadamard gate $H=\frac{X+Z}{\sqrt{2}}$ as well as $\frac{Y+Z}{\sqrt{2}}$ and $\frac{X+Y}{\sqrt{2}}$, all of which are unitary. Given these solutions, one can therefore build an Evenbly code by placing $A'$ tensors with arbitrary orientation on a hyperbolic lattice and then choosing the associated $B$ tensor on each edge.

\begin{figure*}[ht]
\centering
\includegraphics[width=0.95\textwidth]{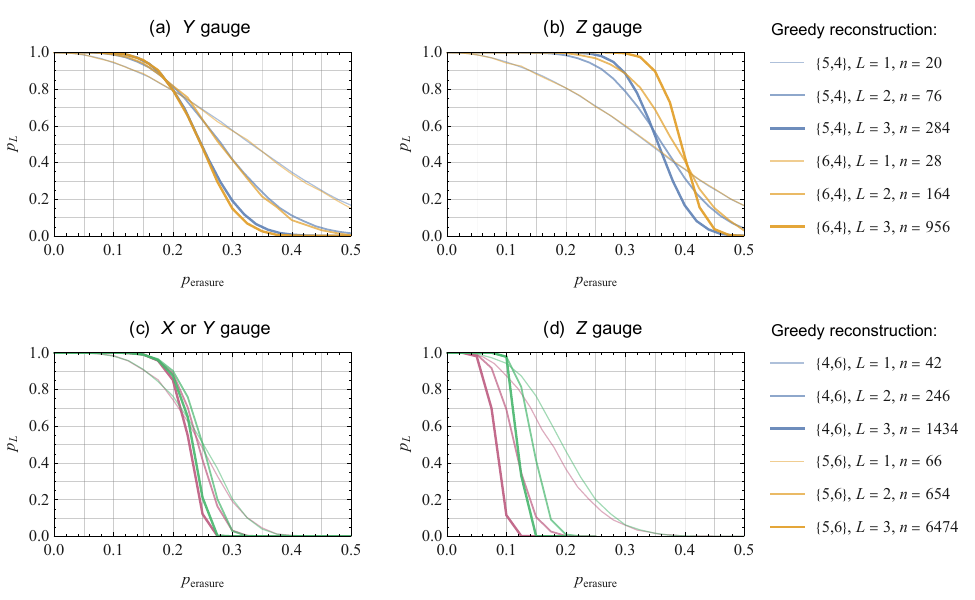}
\caption{Greedy recovery probabilities $p_{L}$ for Evenbly codes on more general $\{p,q\}$ tilings under erasure noise with probability $p_\text{erasure}$. We only consider the zero-rate code with $L$ layers of vertex inflation, with a single logical qubit, under various gauges. (a)-(b): The $q=4$ code considered in the main text can be placed on $\{p,4\}$ tilings for any $p\geq 5$. In the $Y$ gauge, recovery under the greedy algorithm does not appear to depend strongly on $p$, though in the $Z$ gauge the threshold seems to slightly increase with $q$.  
(c)-(d) A similar analysis can be performed for the $q=6$ code for any $p \geq 4$, for which greedy reconstruction under the $X$ gauge becomes non-trivial and equivalent to the $Y$ gauge. Similar to the $q=4$ code, increasing $p$ does not significantly alter $p_L$ at fixed $L$ under $X$/$Y$ gauge but improves it under $Z$ gauge. For these graphs, we have not shown the zeroth layer ($L = 0$) with a single seed tensor, as it exhibits the same recovery curve for any setting (cf.\ \cref{figure:rate_zero_erasure}).
}
\label{figure:greedy_thresholds_extra}
\end{figure*}

\section{Generalizations of Evenbly codes for Other Hyperbolic Tilings}\label{section:appendix_css_other_threshold_data}

The $\{5,4\}$ Evenbly code discussed in the main text can be generalized into a family of infinitely many variants on $\{p,q\}$ tilings for even $p\geq 4$ and $p>\frac{2q}{q-2}$, the latter condition ensuring hyperbolicity.
Using the greedy reconstruction algorithm, we compute recovery curves in \cref{figure:greedy_thresholds_extra} for some of these generalizations under erasure noise, in several different gauge choices; note that results for the ungauged max-rate code as well as the $X$ gauge for $q=4$ are omitted here for brevity. In (a) and (b), we showcase the $p = 5$ and $p = 6$ result for $q = 4$. For these plots, we observe that as we increase $p$, the greedy threshold increases commensurately in the $Z$ gauge; however, as the same procedure is undertaken in the $Y$ gauge, no improvement is seen. Similarly, (c) and (d) convey the results obtained for $q =6$ while increasing $p$ from $5$ to $6$. Again, in the $Y$ gauge, we note that almost no change in the threshold $p_{\text{erasure}}$ occurs. 
Note that greedy reconstruction for $q \geq 6$ is the same in $X$ and $Y$ gauge, as the corresponding greedy steps (Fig.\ \ref{fig:greedy_rec_2}) are identical. The $X$ gauge thus behaves very differently than in the $q=4$ case, as its eigenstates are no longer composed of EPR pairs. Equivalently, in the operator picture applying $\bar{X} = XIXIXI$ for $q=6$ increases the operator weight when pushing $X$ operators to the boundary, while $\bar{X} = XIXI$ for $q=4$ does not.
Finally, the $Z$ gauge of both codes echo the same tendencies found in the $q = 4$ codes: namely, as we increment from $p = 4$ to $p = 5$, we see that the threshold increases for the resulting code. As growing $p$ reduces the rate of the code, it is indeed expected that the central logical-qubit thresholds, as shown, would also improve.  

\section{Alternative Evenbly code Constructions}\label{section:appendix_non_css_htn_codes}

Here, we present several constructions of \emph{alternative} Evenbly codes, using non-CSS constructions of vertex tensors $A'$, as well as the typical $B = H$ edge tensor condition. In addition to the two constructions provided in \cref{section:htn_codes_from_planar_2_uni_states,section:appendix_htn_code_from_a_graph_state}, one may also formulate such alternative Evenbly quantum codes using the complex amplitudes from full 2-uniform states as the $A'$ tensors. Previous work has shown that it is possible to generate new stabilizer quantum error correction codes from old ones by utilizing a procedure known as \emph{code shortening} \cite{gottesman_thesis,raissi_modify_method,macwilliams_sloane_classical_ec}. Here, we show that by employing $(q+1)$-partite 2-uniform states, one can perform logical implantation in the bulk using such 2-uniform quantum states. The implanted 2-uniform state can be then ``shortened" from a $\llbracket q+1,0,k+1\rrbracket$ code to a $\llbracket q,1,k+1\rrbracket$ stabilizer code. We provide an example originating from a $\{5,5\}$ tiling below. 

Consider an HTN ansatz constructed on a $\{5,5\}$ tiling; one may replace the 5-partite 1-uniform GHZ states defined at every vertex with a 6-partite 2-uniform state of the following form: 
\begin{align}
\ket{\Psi_{\llbracket 6,0,3\rrbracket_{2}}} = \frac{1}{\sqrt{8}} \big( &\ket{111111} + \ket{101010} + \ket{001100} \nonumber\\
+ &\ket{011001} + \ket{110000} + \ket{100101} \nonumber\\
 + &\ket{000011} + \ket{010110} \big) \ ,
\end{align}
with stabilizer generators of the form
\begin{subequations}
\begin{align}
\mathcal{G}_{1} &= XXYYZZ, \\
\mathcal{G}_{2} &= XXZZYY, \\
\mathcal{G}_{3} &= XZZXXZ, \\
\mathcal{G}_{4} &= XYYXIZ, \\
\mathcal{G}_{5} &= YXIZXY, \\
\mathcal{G}_{6} &= YYYYII,
\end{align}    
\end{subequations}
which form a $\llbracket 6,0,3\rrbracket$ quantum error correction code. Upon shortening, the code is converted into a $\llbracket 5,1,2\rrbracket$, with new logical operators defined as
\begin{subequations}
\begin{align}
\bar{Z}_{\llbracket 6,0,3\rrbracket} &= XXYYZZ = \mathcal{G}_{1}, \\
\bar{X}_{\llbracket 6,0,3\rrbracket} &= XXZZYY = \mathcal{G}_{2}. 
\end{align}
\end{subequations}
The new stabilizer generators $\{\bar{\mathcal{G}}_{l}\}$, where $l = (n-k)-2 = 4$, are defined as 
\begin{subequations}
\begin{align}
\bar{\mathcal{G}}_{1} &= \mathcal{G}_{3} \circ \bar{Z}_{\llbracket 6,0,3\rrbracket} = IYXZYI, \\ 
\bar{\mathcal{G}}_{2} &= \mathcal{G}_{4} \circ \bar{Z}_{\llbracket 6,0,3\rrbracket} = IZIZZI, \\
\bar{\mathcal{G}}_{3} &= \mathcal{G}_{5} \circ \bar{X}_{\llbracket 6,0,3\rrbracket} = ZIZIZI, \\ 
\bar{\mathcal{G}}_{4} &= \mathcal{G}_{6} = YYYYII.
\end{align}
\end{subequations}
Our new set of generators is now formed simply by removing the last (identity) Pauli from the full set described above, and are given by the stabilizer $\mathcal{S}_{\llbracket 5,1,2\rrbracket} = \{IYXZY,IZIZZ,ZIZIZ,YYYYI\}$; consequently, the new logical operators are $\bar{X}_{\llbracket 5,1,2\rrbracket} = XXZZY$, and $\bar{Z}_{\llbracket 5,1,2\rrbracket} = XXYYZ$. As in \cref{section:htn_codes_from_planar_2_uni_states}, one may apply Hadamard gates to each of the boundary qubits for the stabilizers and logical operators, effectively flipping $X \mapsto Z$, and $Z \mapsto X$, as expected. We also checked numerically that the isometry constraints of \cref{theoremHTNcodes} are upheld, with the only difference being that rotational invariance is not upheld for this 2-uniform state.

There are other constructions of alternative Evenbly codes that exist; we present a few of them below. 

As an example, one can define an Evenbly code for the $\{5,7\}$ hyperbolic tiling, as we can define a vertex tensor $A'$
\begin{align}
\ket{\Psi_{\llbracket 7,0,3\rrbracket_{2}}} = \frac{1}{\sqrt{8}} \big( &\ket{1111111} + \ket{0101010} \nonumber\\ 
 + \ket{1001100} +&\ket{0011001} + \ket{1110000} \\ 
 + \ket{0100101}+&\ket{1000011} + \ket{0010110} \big) \ . \nonumber
\end{align}
This state is 2-uniform, and was taken from \cite{genuine_mme_ortho_arrays}. Here, as in all examples of this section, we utilize stabilizer shortening in order to convert these MME states into codes. In addition to the solutions presented thus far, we have also found other 2-uniform solutions which generate alternative Evenbly codes for hyperbolic $\{p,q\}$ tilings with $q=8,9,10$, respectively:
\begin{widetext}
\begin{subequations}
\begin{align}
\ket{\Psi_{\llbracket 8,0,3\rrbracket_{2}}} = \frac{1}{\sqrt{12}} \big(& \ket{00000000} + \ket{00011101} + \ket{10001110} + \ket{01000111} \nonumber\\ 
&+ \ket{10100011} + \ket{11010001} + \ket{01101000} + \ket{10110100} \\ 
&+ \ket{11011010} + \ket{11101101} + \ket{01110110} + \ket{00111011} \big) \ , \nonumber\\
\ket{\Psi_{\llbracket 9,0,3\rrbracket_{2}}} = \frac{1}{\sqrt{12}} \big(& \ket{000000000} + \ket{100011101} + \ket{010001110} + \ket{101000111} \nonumber\\
&+ \ket{110100011} + \ket{011010001} + \ket{101101000} + \ket{110110100} \\ 
&+ \ket{111011010} + \ket{011101101} + \ket{001110110} + \ket{000111011} \big) \ , \nonumber\\
\ket{\Psi_{\llbracket 10,0,3\rrbracket_{2}}} = \frac{1}{\sqrt{12}} \big(& \ket{0000000000} + \ket{0100011101} + \ket{1010001110} + \ket{1101000111} \nonumber\\ 
&+ \ket{0110100011} + \ket{1011010001} + \ket{1101101000} + \ket{1110110100} \\ 
&+ \ket{0111011010} + \ket{0011101101} + \ket{0001110110} + \ket{1000111011} \big) \ .\nonumber
\end{align}
\end{subequations}
\end{widetext}

Similarly to the constructions of \cite{evenbly_htn_ansatz,conformal_props_steinberg,steinberg_2023}, one may also be able to utilize higher local-dimension k-uniform quantum states in order to design qudit-level Evenbly codes. This direction is the subject of ongoing research. As stated in the main text, decoding these resulting alternative Evenbly codes is highly non-trivial using the greedy recovery algorithm, although in principle it may still be possible. As such, one may utilize the Gaussian elimination algorithm as detailed in \cref{section:appendix_decoder_details}.

\section{Decoder Details}\label{section:appendix_decoder_details}

We utilize in this paper two different decoding schemes against erasure noise. Typically, one describes erasure noise as a specialized form of depolarizing noise in which exact knowledge about the locations of errors is given. Such a channel $\mathcal{E}(\rho)$ acting on density matrix $\rho$ takes on the following form \cite{linear_erasure1,linear_erasure2,css_harris}:
\begin{equation}
\mathcal{E}(\rho) = (1-p)\rho + p/3 \big( \sum_{i}\mathcal{P}_{i}\rho \mathcal{P}_{i} \big),
\end{equation}
where, as we mentioned before, it is assumed that the precise location of errors is known and can be given to a decoder.

\begin{figure*}[p]
    \centering
    \includegraphics[width=0.9\textwidth]{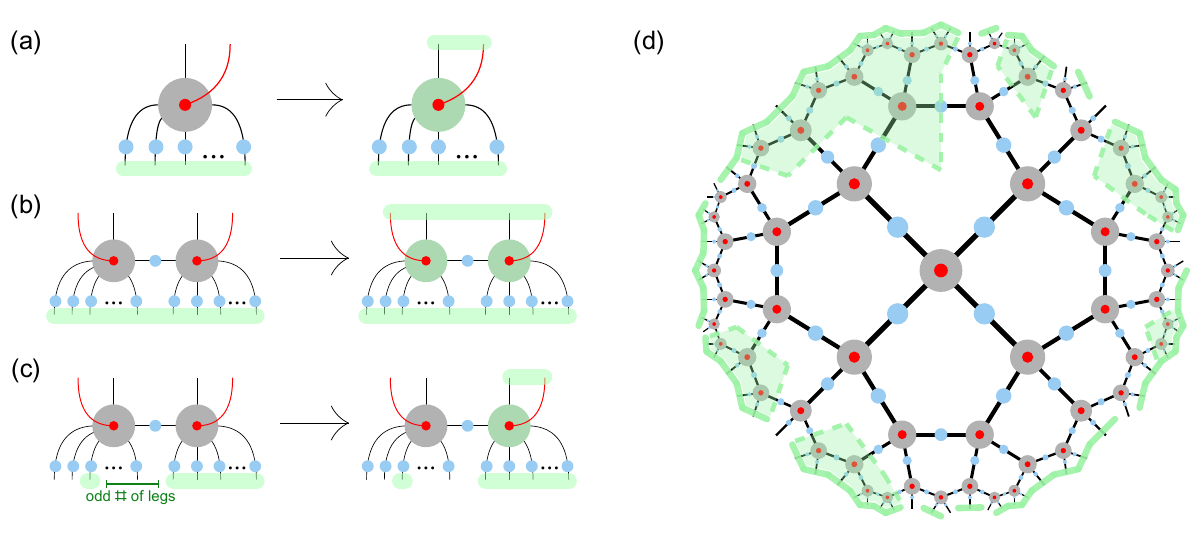}
    \caption{Greedy reconstruction for the max-rate Evenbly code. Given access to a sufficient number of physical sites (green-shaded tensor legs), one can reconstruct operators on the remaining ones. When operators on the logical (red) and all physical legs (black) of a site/vertex are recoverable, we consider the bulk site to be part of the reconstructed \emph{greedy wedge}.
    (a)-(b) Greedy steps equivalent to the Evenbly code's isometry conditions (\cref{fig:HTNcode_isometry_constraints}).
    (c) An additional greedy step that is possible for the GHZ-based Evenbly code defined by \cref{eq:EQ_XZCODE_STAB}.
    (d) An example of the greedy wedge (shaded region) after iteratively applying the steps (a)-(c), starting from a subset of physical boundary sites (thick boundary lines) and terminating at a ``Ryu-Takayanagi'' cut  (dashed bulk lines). Note that loss of access to only a few boundary sites restricts reconstruction deeper in the bulk.
    }
    \label{fig:greedy_rec_1}
\end{figure*}

\begin{figure*}[p]
    \centering
    \includegraphics[width=0.9\textwidth]{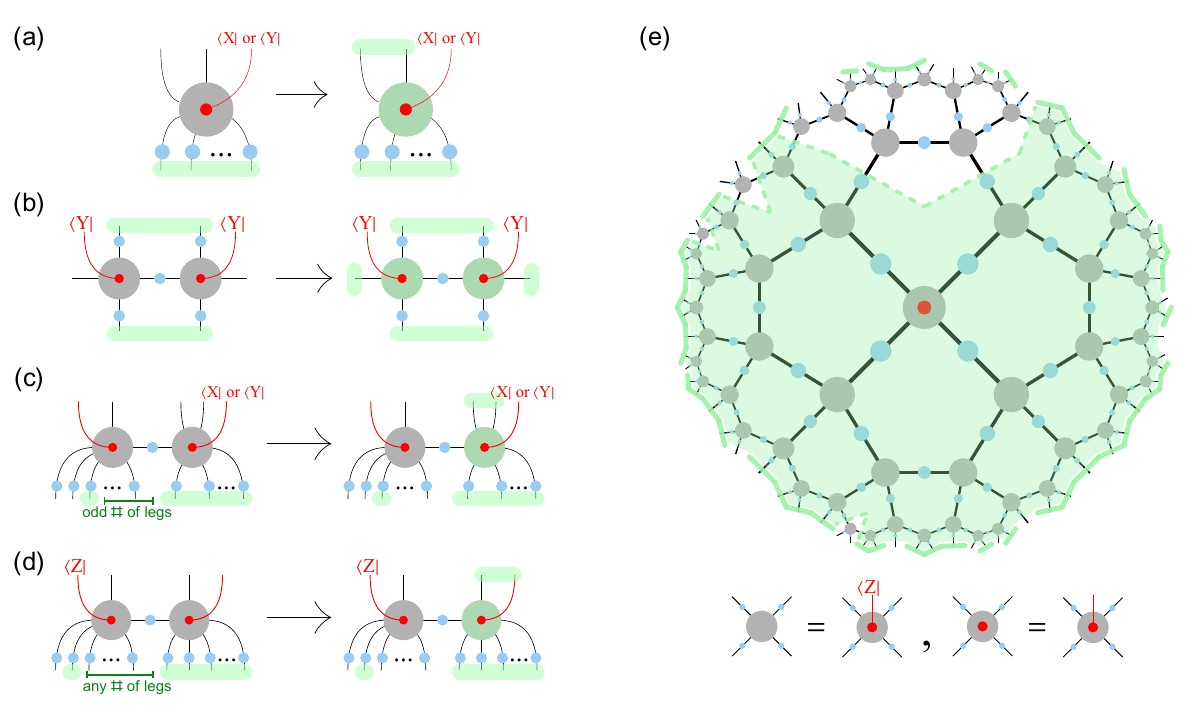}
    \caption{Greedy reconstruction for the zero-rate GHZ-based Evenbly code. (a)-(d) Greedy steps for the Evenbly code under gauging. Note that step (b) only applies to the $q=4$ code. (e) An example of the greedy wedge (shaded region) under $Z$ gauge, iteratively applying the step (d) in addition to those in \cref{fig:greedy_rec_1}(a)-(c). Under this gauge, reconstruction of the central bulk qubit is possible even when access to a sparse number of boundary sites is lost.
    }
    \label{fig:greedy_rec_2}
\end{figure*}

\subsection{Greedy Reconstruction Algorithm} \label{section:appendix_greedy}

A non-optimal but efficient decoder against erasure errors on holographic codes is given by a greedy algorithm, first proposed in \cite{happy_code} for HaPPY codes. In holographic codes, the logical qubits are arranged on a hyperbolic lattice, resulting in a hierarchy with respect to bulk depth, i.e., lattice distance from the physical qubits on the boundary. After some number of known (heralded) erasures of the physical qubits has occured, one starts to reconstruct the logical qubits closest to the boundary, applying the decoding map corresponding to each individual tensor (e.g., of the $\llbracket 5,1,3\rrbracket$ code for the perfect-tensor pentagon code). Once these are (partially) reconstructed, one has access to the ``intermediate'' physical qubits corresponding to contracted legs one layer deeper into the bulk, repeating the local reconstruction procedure there then repeating the procedure. When no more logical qubits can be reconstructed, the greedy algorithm terminates and one has effectively found a decoder for logical qubits in a \emph{greedy wedge} of the bulk on the non-erased physical qubits. 

A greedy decoder can be straightforwardly constructed for any tensor network stabilizer code by translating possible operator pushing steps, which follow from the application of the stabilizers and gauge-fixed logical operators, into greedy steps. For any Evenbly code, for example, the operator pushing steps of \cref{fig:XZCODE_OPREC} (which are equivalent to the generalized isometry conditions of \cref{fig:HTNcode_isometry_constraints}) imply a greedy recovery for single and bulk-neighboring logical qubits that is shown in \cref{fig:greedy_rec_1}(a)-(b). Note that this generalizes the original greedy algorithm of \cite{happy_code}, in which each greedy step only recovers one logical qubit at a time.

Beyond these general recovery step that any Evenbly code fulfills, the GHZ-based code discussed in the main text allows for an additional greedy step shown in \cref{fig:greedy_rec_1}(c), whereby access to physical sites on a tensor can assist in the recovery of the logical qubit on its neighbor.

In spite of these limiting factors, greedy recovery for a max-rate Evenbly code is generally much weaker than for a HaPPY code, owing to the effect that low-weight errors can have deep in the bulk \cite{steinberg_2023}, precluding the HaPPY codes' feature of ``uberholography'' \cite{uberholography}. A generic example is given in \cref{fig:greedy_rec_1}(d), where the erasure of only a few boundary sites prevents greedy reconstruction of most of the bulk.

Fortunately, the greedy algorithm is significantly strengthened in the gauge-fixed regime, where the availability of additional stabilizer terms on each tensor allows for a wider range of greedy reconstruction steps. In the GHZ-based Evenbly code, some greedy steps following gauge-fixing in the X, Y, or $Z$ gauge are visualized in \cref{fig:greedy_rec_2}(a)-(d). Crucially, some of these steps extend the greedy wedge from non-adjacent sites, allowing disjoint wedges to merge in the bulk and preventing small boundary erasures from affecting recovery deep in the bulk. Again we give a generic example of such a reconstruction in \cref{fig:greedy_rec_2}(e).

The greedy algorithm is not an optimal erasure decoder, as it only allows for reconstruction of a logical operator if an entire bulk wedge containing the relevant logical sites can be reconstructed as well. However, it is very fast: Given $k$ logical qubits, each greedy step involves checking every bulk site that is not yet in the greedy wedge (that is, at most $k$). As the algorithm terminates if a greedy step does not add at least one more logical qubit to the wedge, the number of steps is likewise bounded by $k$, resulting in a complexity of $O(k^2)$.

\subsection{Gaussian Elimination Algorithm} \label{section:appendix_gaussian_elim}

The Gaussian elimination algorithm, as introduced in \cite{css_harris,harris_phd_thesis}, functions as follows. Firstly, the stabilizers $\mathcal{S}_{R}$ and logical operators $\mathcal{L}_{R}$ are used as input into the algorithm, along with the number of physical qubits $n$. Next, the stabilizers and logical operators are converted into \emph{binary symplectic vectors} \cite{lidar_qec_book,gottesman_bsv}; each vector is arranged as an $((n-k) \times n)$ array in which the last two rows represent the binary symplectic support of the two logical operators (in the case of the zero-rate Evenbly code), and we name these $BSV$ in Algorithm \ref{alg:gaussian_elimination_alg}. For every Monte Carlo trial, an error is initialized and takes the form of an error of weight $w = [0, n]$. As an example, a $w = 1$ error on the $L = 0$ HTN seed code could be randomly initialized in BSV form as 

\begin{equation}
\mathbf{e} = [1000|1000] \ .
\end{equation}

After initializing a randomized error of weight $w$, we perform stabilizer and logical operator \emph{filtering}, equivalent to the treatment in \cite{css_harris}. After obtaining the filtered-support array $BSV'$ from $\text{BSV\_filter}$, we perform Gaussian elimination modulo 2; the matrix row-reduction algorithm that we utilize exhibits runtime complexity as $\mathcal{O}((n-k)^{2}n)$, although faster methods that guarantee optimality could be leveraged \cite{optimality_gaussian_alg}. 

After obtaining the binary reduced-row echelon form (RREF) $BSV'_{\text{RREF}}$, we iterate through the rows of $BSV'_{\text{RREF}}$ and check the last two columns. If either of the entries are of value 1, then we check the other values in the other $(n-2)$ columns. If any value is equal to 1, then the algorithm succeeds, and we return $\mathrm{True}$ after incrementing the counter $\text{num\_true}$. The algorithm terminates when all trials have completed, and the number of $\mathrm{True}$ values are counted against the number of trials (which we name $\mathcal{P}_{\text{rec}} = \frac{\text{num\_true}}{\text{mc\_trials}}$). This mathematical procedure largely follows \cite{css_harris}. 

\begin{table}[t]
\centering
{\small
\begin{tabular}{|c | c | c | c|} 
 \hline
Layer & HTN & 1 Extra Layer & 2 Extra Layers \\ [0.5ex] 
 \hline
L = 0 & $1.0e 5$ & $1.0e 5$ & $1.0e 5$ \\ 
L = 1 & $1.0e 5$ & $1.0e 5$ & $1.0e 5$ \\
L = 2 & $1.0e 4$ & $1.0e 4$ & $1.0e 4$ \\ 
L = 3 & $1.0e 3$ & $1.0e 3$ & $1.0e 3$ \\ 
L = 4 & $1.0e 3$ & $1.0e 3$ & $1.0e 3$ \\ [1ex] 
 \hline
\end{tabular}
}
\caption{Monte Carlo trial data for the simulations carried out in this study with the Gaussian elimination and integer-optimization decoders.}
\label{table:monte_carlo_trial_data}
\end{table}

Finally, in order to obtain estimates for the error recovery probability, we use the binomial formula 
\begin{equation}
P_{\text{recovery}}(p,n) = \sum_{w} {n \choose w} p^{w} (1-p)^{n-w} \mathcal{P}_{\text{rec}} \ .  
\end{equation}
In the simulations executed for this work, we vary the number of Monte Carlo simulations as a function of the number of layers, in order to efficiently delegate computational resources. The number of Monte Carlo trials per layer and variant tested are shown in \cref{table:monte_carlo_trial_data}. The first column, labeled "HTN", corresponds to the zero-rate and constant-rate Evenbly codes tested in \cref{section:erasure_threshold,section:constant_rate}, while the others refer to the other variants of the constant-rate HTN in which extra layers of Z or $Y$ gauge are added.

\begin{algorithm}[htb]
\small
\caption{Pseudocode for the Gaussian elimination algorithm. We define all of the terms related in this algorithm in \cref{section:appendix_gaussian_elim}.}
\label{alg:gaussian_elimination_alg}

\KwIn{$\mathcal{S}_{R}, \mathcal{L}_{R}, n, \text{mc\_trials}$}

$\text{BSV} \gets \text{convert\_to\_BSV}(\mathcal{S}_{R}, \mathcal{L}_{R})$ \\
$\text{num\_true} = 0$ \\
\For{$\text{trial} \in \text{mc\_trials}$}{
\For{$w \in [ 0 , n ]$}{
$\mathbf{e} \gets \text{initialize\_error\_vector}(w,n)$ \\
$BSV' \gets \text{BSV\_filter}(\mathbf{e},BSV)$ \\
$BSV'_{RREF} \gets \text{gaussian\_elim}(BSV')$ \\
\For{$\text{row} \in \text{rows}$}{
\eIf{$BSV'[\text{row}][-1] \text{ or } BSV'[\text{row}][-2] == 1$}{
\If{$\text(any(BSV'[\text{row}][:-2]) == 1$}{
\textbf{continue}
}
}{
\textbf{return} False
}}
$\text{num\_true} = \text{num\_true} + 1$ \\
\textbf{return} True
}}
\textbf{return} $\mathcal{P}_{\text{rec}} = \text{num\_true} / \text{mc\_trials}$
\end{algorithm}

\begin{algorithm}[htb]
\small
\caption{Pseudocode for one iteration of the integer-optimization decoding algorithm.}\label{alg:int_opt_decode_alg}
\KwIn{$\mathcal{S}_{R}, \mathcal{L}_{R}, n$} 
$\mathcal{S} \gets \text{convert\_to\_BSV}(\mathcal{S}_{R})$ \\
$\epsilon \gets \text{initialize\_error\_vector}(w,n)$ \\
$\mathcal{S}_{\text{pseudo}}^{-1} \gets \text{find\_pseudoinverse\_matrix}(\mathcal{S})$ \\
$\epsilon' \gets \text{minimize\_pauli\_weight}(\mathcal{S}_{\text{pseudo}}^{-1})$ \\
\textbf{return} $\epsilon'$
\end{algorithm}

\subsection{Integer-Optimization Decoding Algorithm} \label{section:appendix_depo_noise_details}

Our approach to the integer-optimization decoder largely follows the treatment of \cite{harris_int_decoder,junyu_msc_thesis}, and proceeds as follows. After initializing all of the stabilizers in BSV form (as discussed in \cref{section:appendix_gaussian_elim}) and stacking them into a $((n-k) \times n)$ matrix $\mathcal{S}$, and calculate the error syndromes $s = \mathcal{S} \cdot \epsilon$, where $\epsilon$ represents a Pauli error initialized of a certain weight $w$. However, as it is known that the matrix $\mathcal{S}$ is not a square matrix, it does not, strictly speaking, have a modulo $2$ inverse. As such, we must seek the \emph{pseudoinverse} $\mathcal{S}_{\text{pseudo}}^{-1}$ for the error $\epsilon$ such that the $\mathcal{S} \cdot \mathcal{S}_{\text{pseudo}}^{-1} = \mathbb{I}$. This matrix $\mathcal{S}_{\text{pseudo}}^{-1}$ exhibits the property that every column anticommutes with the corresponding stabilizer from $\mathcal{S}$. More information on this procedure can be found in \cite{junyu_msc_thesis}. 

Finally, after obtaining the best guess for the error based on the pseudoinverse, we recognize that the suggested error is not of the lowest possible weight; therefore, a second portion of the algorithm involves searching for a minimal-weight logical-equivalent error, solving the combinatorial problem 
\begin{equation}
\epsilon' = \epsilon + \sum^{(n-k)}_{\ell} \lambda_{\ell} s_{\ell} + \sum^{k}_{m} \mu_{m} L_{m},
\end{equation}
where $\lambda_{\ell}, \mu_{m} \in \mathbb{Z}_{2}$. For this, we leveraged the integer-optimization functionality of the Gurobi optimization package \cite{gurobi}.

Judging the correctness of the combinatorial optimization equation above proceeds similarly to the Gaussian elimination algorithm described in \cref{section:appendix_gaussian_elim}; if the algorithm succeeds in finding a minimal-weight vector $\epsilon'$, then recovery is deemed as successful. After a number of Monte Carlo trials, the same binomial formula from \cref{section:appendix_gaussian_elim} is used to estimate the recovery threshold $p_{\text{rec}}$.

\end{document}